\documentclass[aps,amsmath,twocolumn,floatfix,superscriptaddress,prl,10pt,footinbib]{revtex4-2}

\usepackage{amsmath,amssymb,bm}
\usepackage{here}
\usepackage{graphicx}
\usepackage[italicdiff]{physics}
\usepackage{braket}
\usepackage{setspace}[10]
\usepackage{comment}
\usepackage{color}
\PassOptionsToPackage{hyphens}{url}
\usepackage{hyperref}
\usepackage{mathtools}
\usepackage{mathrsfs}
\usepackage{siunitx}

\hypersetup{
	setpagesize=false,
	bookmarksnumbered=true,%
	bookmarksopen=true,%
	colorlinks=true,%
	linkcolor=blue,
	citecolor=red,
}

\graphicspath{{fig/}{./fig/}}

\def\equationautorefname~#1\null{\textrm{~(#1)}\null}
\def\figureautorefname~#1\null{ ~#1\null}
\def\tableautorefname~#1\null{ ~#1\null}

\newcommand{\ttau}[1]{\mathrm{T}_{\tau}\qty[#1]}
\newcommand{\bk}{{\bm{k}}}
\newcommand{\br}{{\bm{r}}}
\newcommand{\bp}{{\bm{p}}}
\newcommand{\brw}{^{a}_{\br}}
\newcommand{\brb}{^{b}_{\br}}
\newcommand{\brwp}{^{a}_{\br'}}
\newcommand{\brbp}{^{b}_{\br'}}
\newcommand{\brx}{{\br+\hat{e}_{1}}}
\newcommand{\bry}{{\br+\hat{e}_{2}}}
\newcommand{\brxm}{{\br-\hat{e}_{1}}}
\newcommand{\brym}{{\br-\hat{e}_{2}}}

\newcommand{\sumk}{\sideset{}{'}\sum_{\bk}}

\newcommand{\tild}{\tilde{d}}

\newcommand{\dc}{{\mathrm{dc}}}
\newcommand{\ac}{{\mathrm{ac}}}

\newcommand{\etams}{\eta_{\mathrm{ms}}}

\newcommand{\edc}{\mathcal{E}_\dc}
\newcommand{\eac}{\mathcal{E}_\ac}

\newcommand{\hms}{H_{\mathrm{ms}}}
\newcommand{\gz}{\mathcal{G}^{(0)}}

\begin{document}
\title{Linear and Nonlinear Optical Responses in Kitaev Spin Liquids}
\author{Minoru Kanega}
\email{20nm010t@vc.ibaraki.ac.jp}
\affiliation{Department of Physics, Ibaraki University, Mito,
	Ibaraki 310-8512, Japan}
\author{Tatsuhiko N. Ikeda}
\email{tikeda@issp.u-tokyo.ac.jp}
\affiliation{Institute for Solid State Physics, University of Tokyo, Kashiwa, Chiba 277-8581, Japan}
\author{Masahiro Sato}
\email{masahiro.sato.phys@vc.ibaraki.ac.jp}
\affiliation{Department of Physics, Ibaraki University, Mito,
	Ibaraki 310-8512, Japan}

\date{\today}
\begin{abstract}
	We theoretically study THz-light-driven high-harmonic generation (HHG) in the spin-liquid states
	of the Kitaev honeycomb model with a magnetostriction coupling between spin and electric polarization.
	To compute the HHG spectra, we numerically solve the Lindblad equation, taking account of the dissipation effect.
	We find that isotropic Kitaev models possess a dynamical symmetry, which is broken
	by a static electric field, analogous to HHG in electron systems.
	We show that the HHG spectra exhibit characteristic continua of Majorana fermion excitations,
	and their broad peaks can be controlled by applying static electric or magnetic fields.
	In particular, the magnetic-field dependence of the HHG spectra drastically differs from those of usual ordered magnets.
	These results indicate that an intense THz laser provides a powerful tool to observe dynamic features of quantum spin liquids.
\end{abstract}
\maketitle


\textit{Introduction.}--
Quantum spin liquids (QSLs) have attracted tremendous attention for decades as exotic states of matter.
Many theoreticians have tried to find essential properties of QSLs,
and it has been theoretically revealed that the low-energy excitations of QSLs are given by
fractionalized particles and the wave functions possess a topological nature~\cite{Wen2007,Balents2010,Savary2017,Knolle2019}.

Meanwhile, it has been recognized as notoriously difficult to identify QSLs experimentally
because most of their thermodynamic quantities are featureless.
Hence, the experiments have been done for their dynamical quantities.
For instance, longitudinal~\cite{Yamashita2010,Sologubenko2001,Hirobe2017}
and transverse~\cite{Onose2010,Watanabe2016} transport phenomena
have provided important information about low-energy excitations in QSLs.
For the Kitaev QSL~\cite{Trebst2017,Hermanns2018,Takagi2019,Motome2020a,Motome2020}, ferromagnetic $\alpha$--$\mathrm{RuCl_3}$ has been shown to exhibit
several characteristic behaviors in, e.g., the thermal Hall effect~\cite{Kasahara2018},
longitudinal thermal conductivity~\cite{Hirobe2017}, and Raman scattering~\cite{Sandilands2015,Nasu2016},
and antiferromagnetic $\mathrm{YbCl_3}$~\cite{Xing2020} has been expected to host QSL
from neutron diffraction measurements.
While these experimental results have been reasonably taken as evidence for QSLs,
they are indirect, and active studies are ongoing to search for new ways to obtain further evidence.

One such direction is the nonlinear optical response at the THz frequency regime, which has been opened up
by rapid development of THz laser technology~\cite{Hirori2011,Sato2013,Dhillon2017,Liu2017}.
Being at the energy scale of magnetic excitations, THz pulses are suitable for directly investigating
and controlling quantum spin systems~\cite{Pimenov2006,Takahashi2012,Kubacka2014,Kezsmarki2014,Mukai2016,Baierl2016,Lu2017,
	Nemec2018,Sirenko2019,Miyahara2008,Mochizuki2010,Mochizuki2010a,Miyahara2012,
	Takayoshi2014,Takayoshi2014a,Sato2014,Sato2016,Fujita2017a,Ikeda2019,
	Ishizuka2019,Ishizuka2019a,Higashikawa2018,Takayoshi2019a,Sato2020}.
To detect crisp signatures of QSLs, the so-called THz two-dimensional coherent spectroscopy
has been proposed~\cite{Wan2019} and theoretically analyzed in the Kitaev model~\cite{Choi2019}.
However, in the Kitaev model, this method is based on third-order,
rather than second-order, optical response and thus it requires much stronger THz pulses for successful detection.
The required intensity of THz pulses can be the bottleneck in experiments
since spin-light couplings are generally much weaker than charge-light ones~\cite{Takayoshi2014a,Sato2014,Sato2016,Fujita2017a,Ikeda2019,Ishizuka2019,Ishizuka2019a}.
In fact, while high-harmonic generation (HHG)~\cite{Ghimire2011,Schubert2014} has been observed
at THz frequencies in, e.g., Dirac electrons~\cite{Hafez2018,Cheng2020,Kovalev2020},
only the second-order response has been reported~\cite{Baierl2016} in magnetic insulators at present.
Thus, another method based on lower-order nonlinear responses, if exists,
should be useful for experimental verification of QSLs.

In this Letter, we show that a combination of an intense THz laser pulse and static electromagnetic fields
uncovers characteristics of the Kitaev QSL through harmonic generation including the second-order harmonic.
We numerically analyze the HHG spectra of the Kitaev model with magnetostriction-type
magnetoelectric (ME) coupling~\cite{Tokura2014} with the quantum master equation approach~\cite{Breuer2007,Alicki2007,Ikeda2019,Sato2020} to take account of dissipation effects.
In addition to broad and continuous response functions characteristic of Majorana fermions,
we find that static electric fields break some symmetry and
activate the second-harmonic generation (SHG), and a static magnetic field causes
an anomalous shift for the harmonic spectra.
These findings indicate that nonlinear response to intense THz light gives us
a powerful instrument for detecting dynamical features of Kitaev QSLs.
Through this study, we will build a bridge between photoscience and QSLs.

\textit{Kitaev Model and Methods.}--
\begin{figure}[tb]
	\centering
	\includegraphics[width=\linewidth]{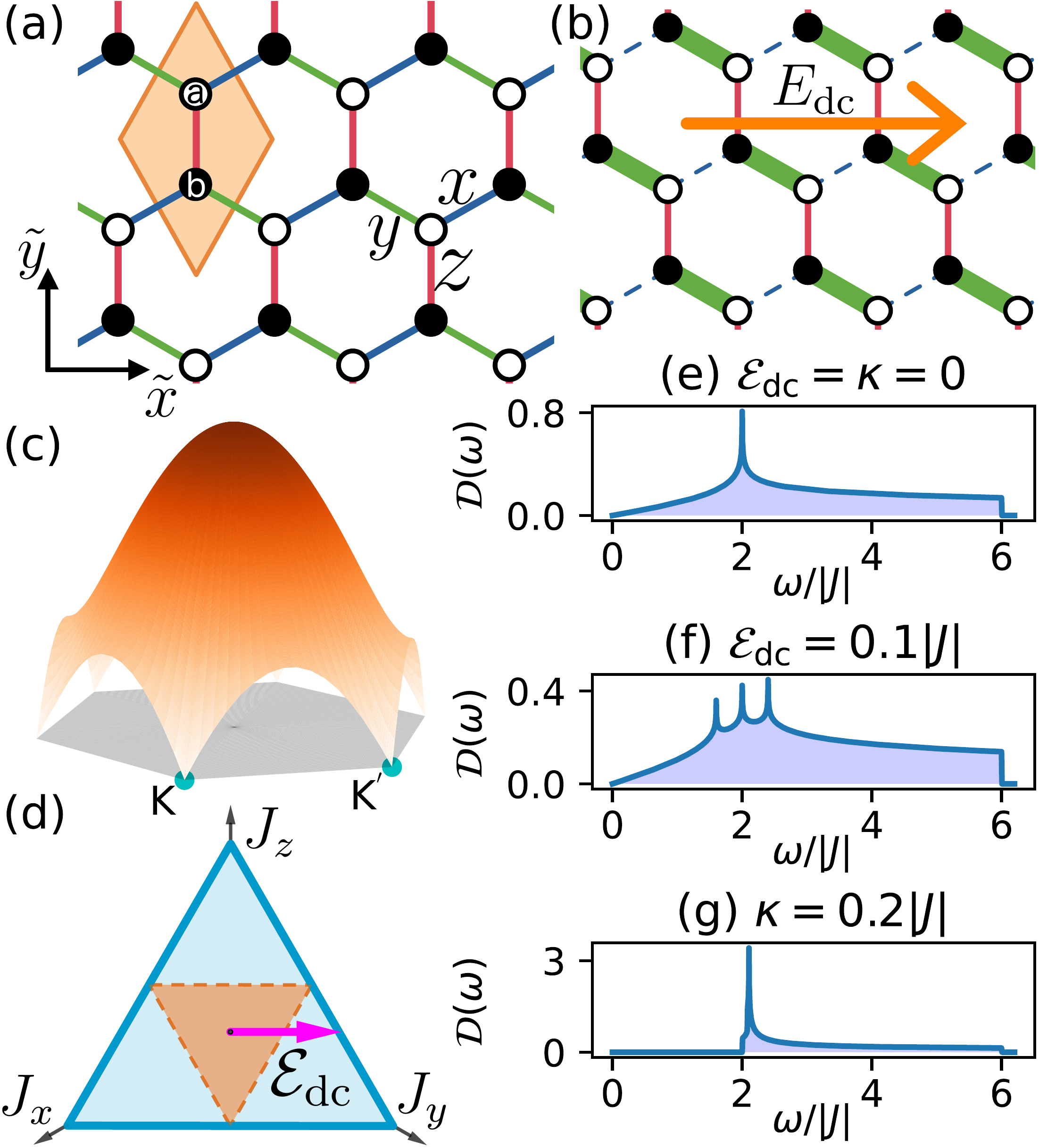}
	\caption{(a) Lattice structure of the Kitaev model. Blue, green, and red lines respectively correspond to $x$, $y$, and $z$ bonds.
		(b) Kitaev model with a dimerization on $x$ and $y$ bonds, which is caused by a static electric field $E_\dc$
		along the $\tilde{x}$ direction.
		(c) Gapless itinerant fermion band of the Kitaev model at $J_{x,y,z}=J$ and $E_\dc=\kappa=0$.
		(d) Ground-state phase diagram of the Kitaev model in $(J_x,J_y,J_z)$ space at $\kappa=0$.
		Orange and blue areas are respectively gapless and gapped QSLs. Application of $E_\dc$ induces
		$(J_x,J_y)\rightarrow(J_x-\edc,J_y+\edc)$.
		(e--g) Density of states of itinerant fermions at $(\edc,\kappa)=(0,0)$, $(0.1,0)$, and $(0,0.2)$ in $J<0$.
		For detail of the density of states, see Figs.~\hyperref[fig:DoS-kappa]{S8} and~\hyperref[fig:DoS-kappa-high]{S9} of Supplemental Material~\cite{Supplemental}.}
	\label{fig:band}
\end{figure}
The Hamiltonian of the Kitaev model (see Fig.\autoref{fig:band})~\cite{Kitaev2006a} for this work is given by
\begin{align}
	\hat{H}_0= & -\sum_{\alpha,\braket{\br,\br'}_\alpha}J_\alpha\sigma^\alpha_{\br}\sigma^\alpha_{\br'}-\kappa\sum_{\mathrm{NNN}}\sigma^x_{\br}\sigma^y_{\br'}\sigma^z_{\br''}-E_\dc\hat{P} 
	\label{eq:staticHamiltonian}
\end{align}
with
\begin{align}
	\hat{P}=\etams\qty\Big(-\sum_{\braket{\br,\br'}_x}
	\sigma^x_{\br}\sigma^x_{\br'}+\sum_{\braket{\br,\br'}_y}\sigma^y_{\br}\sigma^y_{\br'}),
	\label{eq:polarization}
\end{align}
where
$\sigma^\alpha_{\br}$~($\alpha=x,y,z$) is the Pauli matrix on the site $\br=(j,k)$, and
the spin-$1/2$ operator on $\br$ is $S^\alpha_{\br}=\frac{\hbar}{2}\sigma^\alpha_{\br}$
(We set $\hbar=1$ throughout the paper).
$J_\alpha$ is the Ising coupling constant
between nearest neighboring spins on $\alpha$ bond $\braket{\br,\br'}_{\alpha}$
and we mainly focus on the symmetric point $J=J_{x,y,z}$.
The neighboring three spin term~\cite{Kitaev2006a}
originates from the third-order perturbation with respect to a static Zeeman term
$\mathcal{H}_B=-g\mu_B\bm{B}\cdot\sum_{\br}\bm{\sigma}^\alpha_{\br}$ of an applied magnetic field
$\bm{B}$ ($g$ is the g-factor and $\mu_B$ is the Bohr magneton).
The coupling constant is computed as $\kappa\sim(g\mu_B)^3B_xB_yB_z/J^2$:
For $\abs{J}/k_B=\SI{10}{K}$,
$\kappa\sim0.1\abs{J}$ corresponds to $\abs{\bm{B}}\sim\SI{1}{T}$.
The $\kappa$ term is the leading term changing the Majorana-fermion dispersion (as one will see later)
in the Zeeman interaction.
The final term of Eq.~(\ref{eq:staticHamiltonian}) represents the coupling
between electric polarization $\hat{P}$ and an applied static electric field $E_\dc$ along the $\tilde{x}$ direction
(Figs.\autoref{fig:band}(b) and\autoref{fig:band}(d)).
Equation~(\ref{eq:polarization}) assumes that $\hat{P}$ is proportional to dimerization
along the $\tilde{x}$ direction.
This kind of ME terms appears in a class of multiferroic magnets and
its typical origin is the spin-phonon coupling~\cite{Tokura2014}.
This dimerization makes the ground-state energy reduce (see the Supplemental Material~\cite{Supplemental}),
and therefore this ME term may appear in a sort of real Kitaev-like materials.
In a class of multiferroics, the ME-coupling energy reaches that of the
Zeeman term~\cite{Pimenov2006,Miyahara2008,Mochizuki2010a,Takahashi2012,Tokura2014,Kubacka2014,Huvonen2009,Furukawa2010},
and thereby we have assumed that the effective coupling $\edc=E_\dc\etams$
is the same order as $g_0\mu_BE_\dc/c$, with $c$ being the speed of light and $g_0=2$:
For $\abs{J}/k_B=\SI{10}{K}$,
$\mathcal{E}_\dc\sim$ 0.01$\abs{J}$--0.1$\abs{J}$ corresponds to 0.1--\SI{1}{MV/cm}.
We note that a similar dimerization ($J_x\neq J_y$) can appear
in a class of Kitaev candidates even without a dc electric field.

The Kitaev model (\ref{eq:staticHamiltonian}) is exactly solvable via
fermionization~\cite{Chen2008,Kitaev2006a}, as detailed in the literature~\cite{Chen2008,Kitaev2006a,Trebst2017,Hermanns2018,Takagi2019,Motome2020a}
(see the Supplemental Material~\cite{Supplemental} for details).
The fermionized Hamiltonian consists of four kinds of
Majorana fermions: two dispersive and two localized ones (called visons).
Visons have a small gap~\cite{Kitaev2006a} and are absent~\cite{Lieb1994} in the ground state at zero temperature $T=0$.
The Hamiltonian at $T=0$ is therefore described only by the dispersive fermions.
Figures\autoref{fig:band}(c)--\ref{fig:band}(g) show that the dispersive fermion is gapless around
the isotropic point $J_{x,y,z}=J$, while the $\kappa$ term opens a gap.
We note that some perturbations can also be fermionized~\cite{Takikawa2019,Takikawa2020},
and therefore our analysis below is also applicable to such perturbed Kitaev models~\footnote{
	As discussed in Ref.~\cite{Takikawa2019}, if the Kitaev magnet has both a dc magnetic field and a sort of perturbative term, an additional three spin interaction appears and its coupling constant is possibly proportional to $B$. However, this paper has focused on the system with $\kappa\propto B^3$.}.

To consider HHG in the Kitaev model at $T=0$, we introduce
an ac ME coupling between an ac electric field $E_\ac(t)$ and the polarization $\hat{P}$:
$H_{\mathrm{ms}}(t)=-E_\ac(t)\hat{P}$~\cite{Tokura2014}.
We have assumed that $\bm{E}_\ac\parallel\tilde{x}$, and the field is a Gaussian pulse
with a THz frequency $\Omega$:
$E_\ac(t)=E_\ac\cos(\Omega{t})f(t)$, and the envelope function is given by
$f(t)=\exp[-2(\ln2)(t^2/t^2_{\mathrm{FWHM}})]$, where $t_{\mathrm{FWHM}}$ is the full width
at half-maximum of the intensity $E_\ac(t)^2$.
We define the ac coupling constant $\eac=E_\ac\eta_{\mathrm{ms}}$ and consider five-cycle pulses
($t_{\mathrm{FWHM}}/T=5$).
We note that even after application of $E_\ac(t)$, the Hamiltonian $H_0+H_{\mathrm{ms}}(t)$ is described
by a bilinear form of the dispersive fermion in wave-vector $\bk$ space.
There is no vison dynamics in our setup that is valid in sufficiently low temperatures.
We also phenomenologically introduce dissipation effects described by the Lindblad equation so as not to break the block-diagonal structure in $\bm{k}$~\cite{Breuer2007,Alicki2007,Ikeda2019,Sato2020} (for details, see the Supplemental Material~\cite{Supplemental}).
The dissipation is $\bm{k}$-independent and characterized by the dissipation rate $\gamma=0.1J$,
which corresponds to the relaxation time $\tau=1/\gamma\sim\SI{7.6}{ps}$ for $J/k_B=\SI{10}{K}$~\cite{Kirilyuk2010,Oshikawa1999,Oshikawa2002,Furuya2015,Beaurepaire1996,Koopmans2000,Lenz2006,Vittoria2010,Mashkovich2019,Tzschaschel2019}.
We suppose that the system is initially $(t\to-\infty)$ in the ground state, and we solve the Lindblad equation
numerically to obtain the THz-driven nonequilibrium dynamics.

The observable of interest is $\hat{P}$, which is the source of HHG in our model.
Time evolution of polarization is given by
$P(t)=\braket{\hat{P}}_t=N^{-1}\sum_{\bk,k_x>0}\Tr[\rho(\bk,t)P_{\bk}]$, where $P_{\bk}$ is $2\times2$ reduced polarization for the subspace with $\bk$,
and $N$ is the total number of unit cells.
Since the electromagnetic radiation is proportional to $d^2P(t)/dt^2$,
the radiation power at frequency $\omega$ is given by
$I(\omega)=|\omega^2P(\omega)|^2$, where $P(\omega)$ is the Fourier transform of $P(t)$~\cite{Jackson1998}.
Since a constant shift of $P(t)$ does not change $I(\omega)$, we will also use $\Delta{P}(t)=P(t)-P(t_{\mathrm{ini}})$.

\textit{Effect of DC Electric Field and Dimerization.}--
\begin{figure}[tb]
	\centering
	\includegraphics[width=\linewidth]{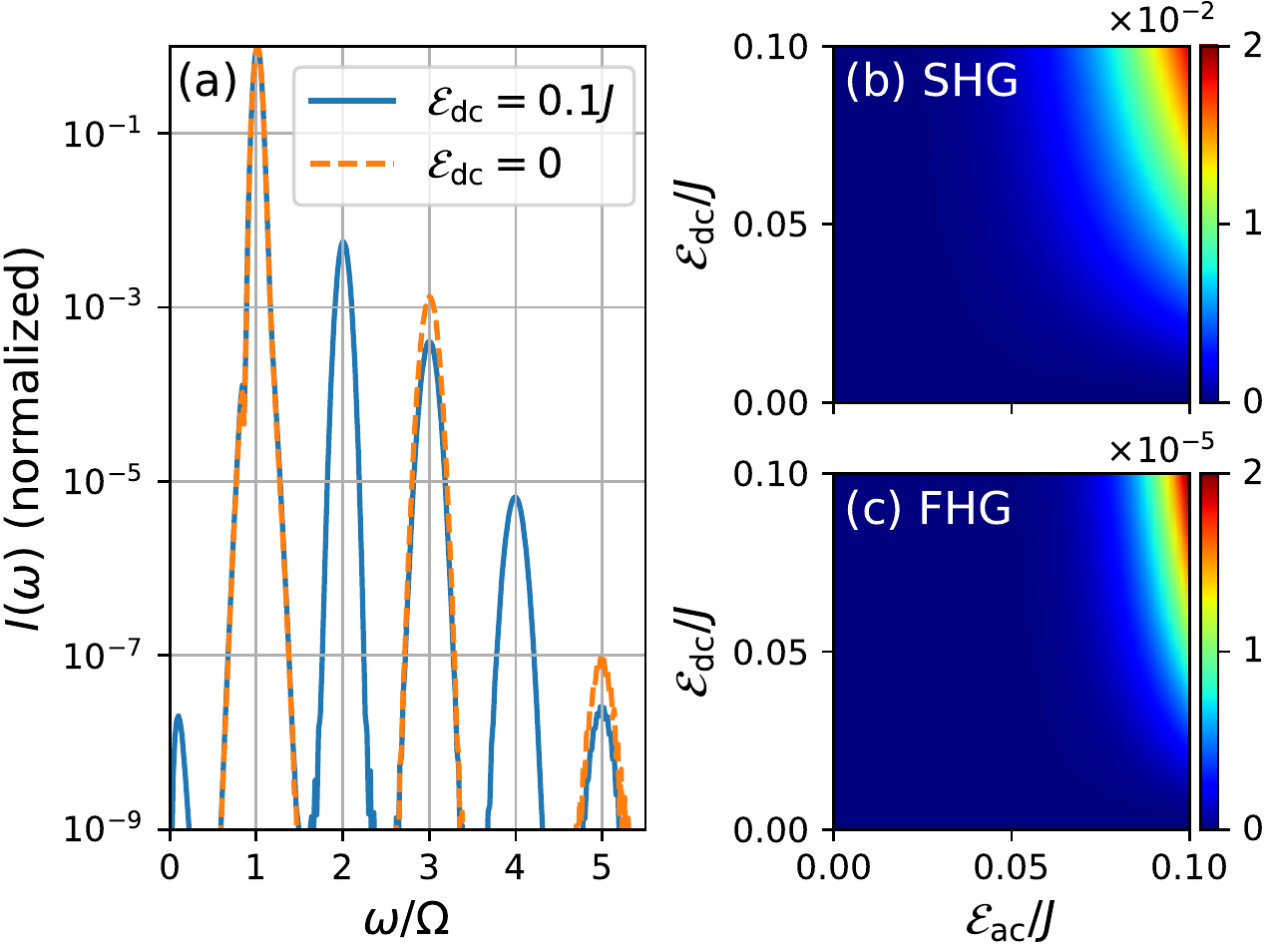}
	\caption{HHG spectra $I(\omega)$ in driven isotropic ($J_{x,y,z}=J$) Kitaev models
		with/without a dc electric field $E_\dc$ at $\kappa=0$ under a THz pulse of $\Omega=2.0J$.
		(a) $I(\omega)$ as a function of $\omega$ at $\edc=0$ and $0.1J$
		under the irradiation of $\eac=0.1J$. $I(\omega)$ is normalized with its maximum value.
		(b), (c) $(E_\ac,E_\dc)$ dependence of SHG [$I(2\Omega)$] and FHG [$I(4\Omega)$] spectra.
		The intensities, panels (b) and (c), are normalized with $I(\Omega)$ at $\eac=\edc=0.05J$.
	}
	\label{fig:Estat-dep}
\end{figure}
We turn to our analyses and results. First we focus on the dc-electric-field dependence of HHG
in the Kitaev model with $J_{x,y,z}=J$ and $\kappa=0$.
Figure\autoref{fig:Estat-dep}(a) shows a typical HHG spectrum $I(\omega)$ for ferromagnetic ($J>0$) Kitaev models at $\edc=0$ and $0.1J$.
Figures\autoref{fig:Estat-dep}(b) and\autoref{fig:Estat-dep}(c) are respectively the $(E_\ac,E_\dc)$ dependence of the SHG and fourth harmonic generation (FHG) that arise due to dimerization by $E_\dc\neq0$.
The SHG signal, activated by the dc electric field, is stronger than higher-order harmonics and can be a useful probe for QSL as we discuss further below.

The HHG selection rules are often understood by dynamical symmetries that become exact in the limit of $t_{\mathrm{FWHM}}\to\infty$~\cite{Alon1998,Neufeld2019} also known as the time glide symmetry~\cite{Morimoto2017}.
For our Kitaev model, we find that the following dynamical symmetry determines the HHG selection rule.
In the non-dimerized case of $E_\dc=0$, the Hamiltonian $H_0+H_{\mathrm{ms}}(t)$
is invariant under the time translation operation $t\to t+T/2$ combined with the unitary transformation
$\hat{U}=\hat{U}_{\mathrm{mir}}\times\hat{U}^{z}_{\pi/2}$, where
$\hat{U}_{\mathrm{mir}}$ is the reflection operation with respect to the $\tilde{y}$ axis
and $\hat{U}^{z}_{\pi/2}$ is the global $\pi/2$ spin rotation around the $S^z$ axis.
$\hat{P}$ is odd for this transformation, and therefore we obtain $P(t+T/2)=-P(t)$,
which means that even-order HHG is prohibited, consistent with Fig.\autoref{fig:Estat-dep}~\cite{Supplemental}.
However, for $E_\dc\neq0$, this dynamical symmetry is broken, and even-order HHG is allowed.
We note that for $\kappa\neq0$, the unitary operator $\hat{U}$ is modified as
$\hat{U}\to\hat{V}\hat{U}$, where $\hat{V}$ is the time-reversal operator.

Thus, even-order HHG can be controlled by the static electric field $E_\dc$ through dimerization.
Similar effects have been discussed in a spin chain~\cite{Ikeda2019} and in electronic systems~\cite{Khurgin1995,Wu2012,Cheng2014,Aktsipetrov2009,Ruzicka2012,Bykov2012,An2013,Moor2017,Nakamura2019,Nakamura2020,Takasan2020}, where $E_\dc$ induces electric current breaking the inversion symmetry.
Being based only on symmetry, the dynamical symmetry argument is applicable to a wide class of perturbed Kitaev models.
For example, the dynamical symmetry survives for $J_z\neq J_x=J_y$ while it breaks down for $J_x\neq J_y$.
The magnetic anisotropy dependence of even-order harmonics is discussed in Supplemental Material~\cite{Supplemental}.
In the following, we assume $E_\dc\neq0$ as necessary to ensure that even-order HHG are present.

\textit{Dependence of Laser Frequency and Intensity.}--
\begin{figure}[tb]
	\centering
	\includegraphics[width=\linewidth]{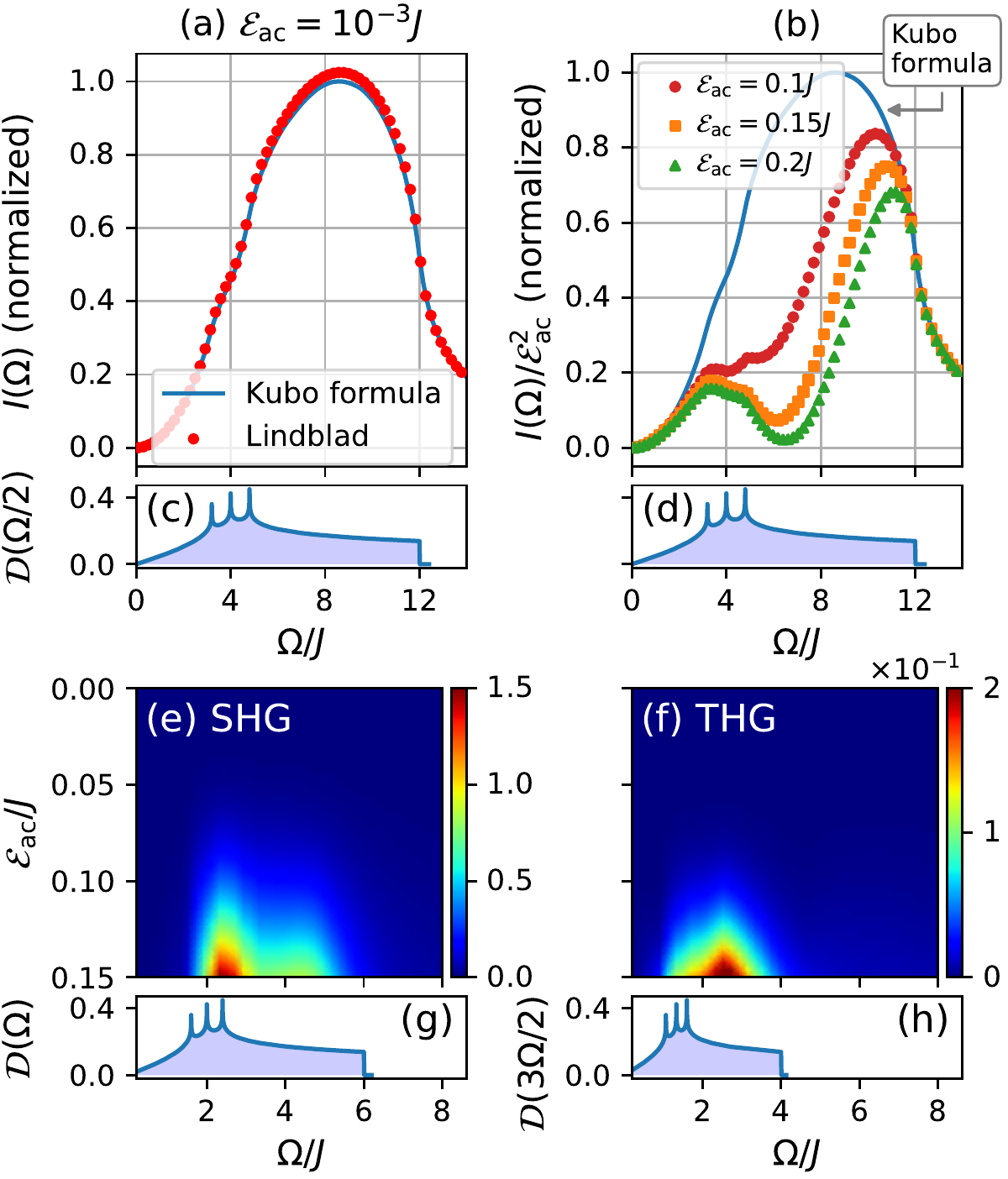}
	\caption{$I(\Omega)$ of the Kitaev models with $J_{x,y,z}=J$ and $\edc=0.1J$ at (a) a
	weak laser intensity $\eac/J=10^{-3}$ and (b) strong intensities of $\eac/J=0.1$, $0.15$, and $0.2$.
	Blue lines and dotted points are respectively results
	of the linear response (Kubo) theory and numerically-solved master equation~\cite{Supplemental}.
	$I(\Omega)$ is normalized with the maximum value of the Kubo formula.
	Panels (c) and (d) show DoSs of fermions corresponding to the cases (a) and (b), respectively.
	(e) SHG [$I(2\Omega)$] and (f) THG [$I(3\Omega)$] of the Kitaev model with $\edc=0.1J$
	in the space $(\eac,\Omega)$.
	The intensities, panels (e) and (f), are normalized with $I(\Omega)$ at $(\eac,\Omega)=(0.05J,2J)$.
	The corresponding DoSs are depicted in panels (g) and (h).}
	\label{fig:omegaExt-dep2d-123HG}
\end{figure}
Next, we consider the $\Omega$ and $E_\ac$ dependence of the HHG.
Figures\autoref{fig:omegaExt-dep2d-123HG}(a) and\autoref{fig:omegaExt-dep2d-123HG}(b), respectively,
show the $\Omega$ dependence of $I(\Omega)$ for weak ($\eac=10^{-3}J$) and
strong ($\eac=0.1J, 0.15J, 0.2J$) THz pulses.
In addition to the numerical result of 
the Lindblad equation,
we plot that of the linear response theory (the Kubo formula)
(see the Supplemental Material~\cite{Supplemental}). 
From the comparison between $I(\Omega)$ in Fig.\autoref{fig:omegaExt-dep2d-123HG}(a)
[Fig.\autoref{fig:omegaExt-dep2d-123HG}(b)] and the fermion density of state (DoS) $\mathcal{D}(\omega)$
in Fig.\autoref{fig:omegaExt-dep2d-123HG}(c) [Fig.\autoref{fig:omegaExt-dep2d-123HG}(d)],
we find that two-particle continuum spectra occur in the driven Kitaev model.
This results from the ME coupling between an ac electric field (i.e., photon) 
and a pair of fermions with $\bk$ and $-\bk$.
This continuum indicates the existence of fermionic excitations in Kitaev magnets, and it differs
qualitatively from usual ordered magnets,
where one often observes a delta-functional peak due to magnons~\footnote{
	Even a single-magnon spectrum can have a broad peak when its lifetime is short. In such a case, it becomes difficult to distinguish the Kitaev continuum and the broad magnon peak. However, combining the HHG experiment with an additional one, we can possibly distinguish these two excitations. For example, magnons always appear in an ordered phase which can be detected by a magnetic Bragg peak, a magnetic phase transition, etc.}.

It is noteworthy that even the fundamental harmonic $(\omega=\Omega)$ shows characteristics of the QSL in the strong THz pulse.
Unlike in the weak pulse, the Kubo formula is no longer valid in the strong one  in the deep nonperturbative regime.
In this regime of $\eac\agt0.15J$, a new broad peak emerges in $I(\Omega)$ at $\Omega_{\mathrm{peak}}\sim4J$ [Fig.\autoref{fig:omegaExt-dep2d-123HG}(b)], which is twice as large as the high DoS position.
Namely, $I(\Omega)$ driven by intense pulses tells us the peak position of the DoS.
For instance, $E_{\mathrm{ac}}=0.15J/\eta_{\mathrm{ms}}\sim\SI{3}{MV/cm}$ for $J=\SI{10}{K}$ under the assumption of
$\eac=g_0\mu_BE_\ac/c$, and it indicates that
currently-available THz laser is strong enough to observe such nonlinear optical spectra.
We note that in Fig.\autoref{fig:omegaExt-dep2d-123HG}(b),
the increase of $I(\Omega)$ in the high-$\Omega$ range around $\Omega\sim8J$
is owing to the factor $\Omega^4$ in $I(\Omega)$.

The SHG and third harmonic generation (THG) spectra,
$I(2\Omega)$ and $I(3\Omega)$, are depicted in Figs.\autoref{fig:omegaExt-dep2d-123HG}(e) and\autoref{fig:omegaExt-dep2d-123HG}(f).
We find that broad peaks in $I(2\Omega)$ and $I(3\Omega)$ appear
around $\Omega=\Omega_{\mathrm{peak}}/2$ and $\Omega_{\mathrm{peak}}/3$, respectively.
This is a natural result, indicating that excitation processes creating fermions with a high DoS
are dominant in HHG.
Figure\autoref{fig:omegaExt-dep2d-123HG}(h) shows that
the peak frequency of the THG is slightly higher than that of the DoS ${\cal D}(3\Omega/2)$.
This would also be attributed to the factor $(3\Omega)^4$ in $I(3\Omega)$.

\textit{Effect of DC Magnetic Field.}--
\begin{figure}[tb]
	\centering
	\includegraphics[width=\linewidth]{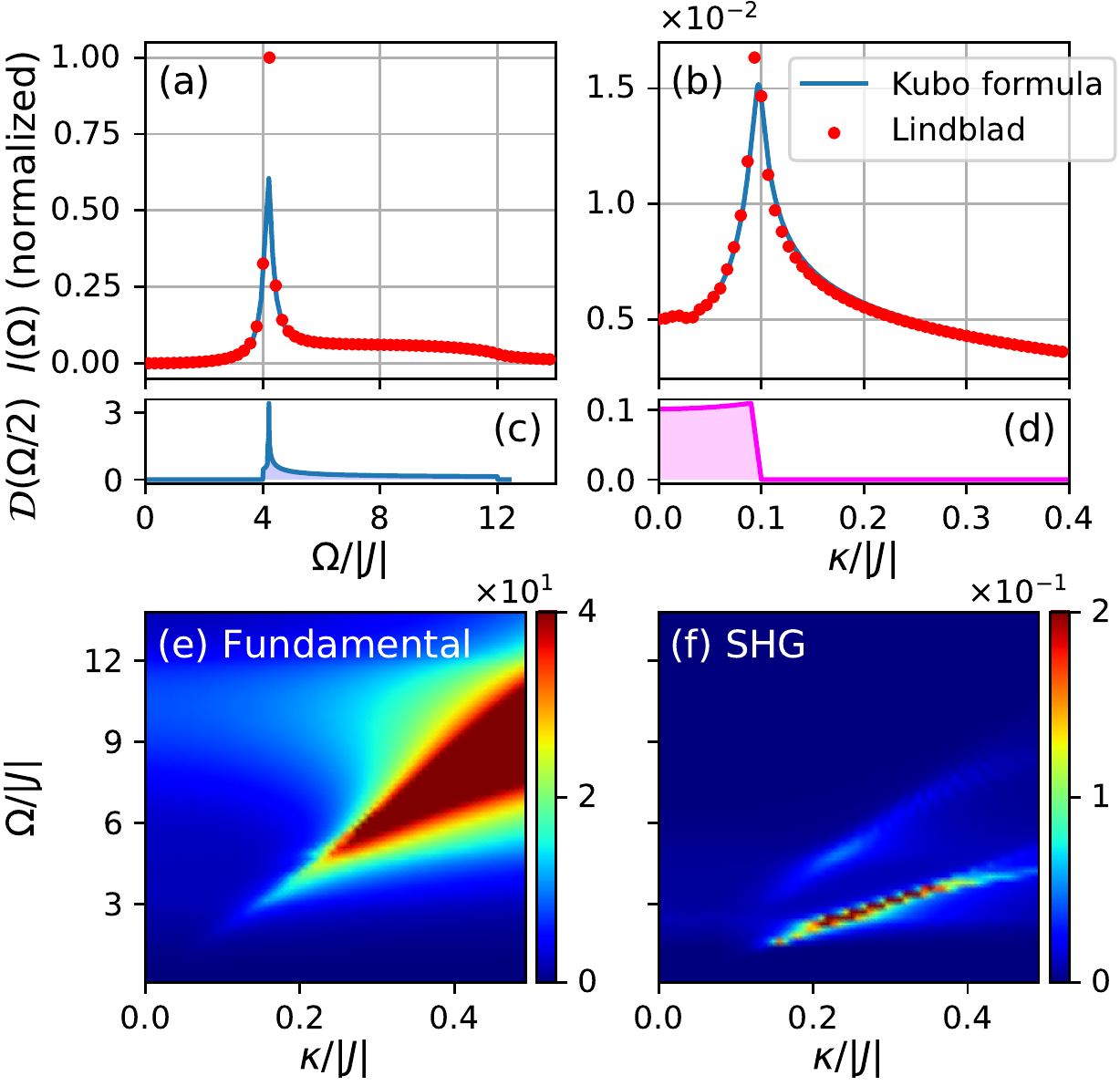}
	\caption{(a) $I(\Omega)$ in an antiferromagnetic ($J<0$) Kitaev model with $\kappa=0.2\abs{J}$, $\edc=0$, and $\eac=10^{-3}\abs{J}$.
	(b) $\kappa$ dependence of $I(\Omega)$ in $\Omega=2.0\abs{J}$, $\edc=0$, and $\eac=10^{-3}\abs{J}$.
	Blue line and dotted points are respectively results of the linear response theory and
	the master equation~\cite{Supplemental}. $I(\Omega)$ in (a) and (b) are normalized
	with the maximum value in the panel (a).
	(c) $\Omega$ and (d) $\kappa$ dependences of $\mathcal{D}(\omega)$ that respectively correspond to panels (a) and (b).
	(e) $I(\Omega)$ and (f) $I(2\Omega)$ of the Kitaev model with $\edc=\eac=0.1\abs{J}$.
	The intensities, panels (e) and (f), are normalized with $I(\Omega)$ at $(\kappa,\Omega)=(0.05\abs{J},2\abs{J})$.
	}
	\label{fig:kappa-dep}
\end{figure}
Now we discuss the dc-magnetic-field dependence of HHG in Kitaev models.
We focus on the antiferromagnetic Kitaev model~\cite{Supplemental}, where the Kitaev QSL state is more stable against magnetic fields than in the ferromagnetic case~\cite{Nasu2018}.
The magnetic-field driven $\kappa$ term opens a mass gap $\Delta_\kappa$ in the fermion band,
as shown in Fig.~\ref{fig:band}(g). As $\Delta_\kappa$ increases,
the maximum of $\mathcal{D}(\omega)$ at $\omega\sim2\abs{J}$ grows up for $\Delta_\kappa\alt2\abs{J}\;(\kappa\alt0.2\abs{J})$ [see Figs.~\ref{fig:band}(e) and~\ref{fig:band}(g)].
Therefore, the intensities of HHG spectra are expected to be controlled
by the dc magnetic field and laser frequency $\Omega$.
Figure\autoref{fig:kappa-dep} proves this expectation.
Panels (a) and (c), respectively, show the $\Omega$ dependence
of $I(\Omega)$ and $\mathcal{D}(\omega)$ at a tuned effective magnetic field $\kappa=0.2\abs{J}$,
where $\mathcal{D}(\Omega)$ take the highest value at $\Omega_0\sim\Delta_\kappa\sim2\abs{J}$.
In this case, we have a sharp peak of $I(\Omega)$ at
$\Omega_{\mathrm{peak}}=2\Omega_0\sim4\abs{J}$.
Figures\autoref{fig:kappa-dep}(b) and\autoref{fig:kappa-dep}(d) represent $I(\Omega)$ and $\mathcal{D}(\omega)$ as a functions of $\kappa$
at a fixed $\Omega=2\abs{J}$. We see that a clear peak of $I(\Omega)$ appears
when $\Delta_\kappa$ passes across one-half of the laser frequency $\Omega/2$.

We also show $I(\Omega)$ and $I(2\Omega)$ in the $(\kappa,\Omega)$ space
in Figs.\autoref{fig:kappa-dep}(e) and\autoref{fig:kappa-dep}(f). For $I(\Omega)$ in the case of $\kappa\agt0.2\abs{J}$,
the frequency $\Omega$ of the broad peak increases monotonically in an almost $\kappa$-linear fashion.
This is because the peak position $\Omega_0$ of $\mathcal{D}(\Omega)$ increases
almost linearly with $\kappa$ for $\kappa\agt0.2\abs{J}$ (for more details, see Fig.~\hyperref[fig:depKappa]{S15} of the Supplemental Material~\cite{Supplemental}).
Since $\kappa\sim\abs{\bm{B}}^3$, the $\bm{B}$-cube-dependent frequency at the peak is specific for the Kitaev model
and essentially differs from the $\bm{B}$-linear behavior of magnon peaks.
The frequency at the peak of $I(2\Omega)$ is almost half of $\Omega_{\mathrm{peak}}$ of $I(\Omega)$,
as shown in Fig.\autoref{fig:kappa-dep}(f), and this is a natural result from the perturbative viewpoint.
We note that,
even in the linear response regime ($\omega=\Omega$ and weak THz pulse),
the fundamental harmonic shows characteristics of the QSL for $\kappa\agt0.2\abs{J}$ (see the Supplemental Material~\cite{Supplemental}).
$I(\Omega)$ driven by intense pulses with a finite magnetic field tells us the peak position of the DoS $\Omega_0$, which is half as large as $\Omega_{\mathrm{peak}}$.

Finally, we estimate the laser intensity required in the HHG experiment.
For the Kitaev magnet with $\abs{J}/k_B=\SI{10}{K}$ and $\kappa=0$,
the required ac electric field $E_\ac$ for the observation of HHG
can be estimated from Fig.\autoref{fig:Estat-dep}(a):
$E_\ac=\SI{2.6}{MV/cm}$ at \SI{0.42}{THz} is necessary for
$R(\Omega)=I(2\Omega)/I(\Omega)\agt10^{-2}$
and $E_\ac=\SI{0.9}{MV/cm}$ for $R(\Omega)\agt10^{-3}$.
These required ac electric fields can be reduced when applying the effective static magnetic field
$\kappa=0.2\abs{J}$
as shown in Fig.\autoref{fig:kappa-dep} and the Supplemental Material~\cite{Supplemental}.
In this case, the electric field is estimated as $E_\ac=\SI{0.7}{MV/cm}$ at \SI{0.42}{THz} for
$R(\Omega)\agt10^{-2}$ and $E_\ac=\SI{0.2}{MV/cm}$ for $R(\Omega)\agt10^{-3}$.
We remark that the above criteria, $R(\Omega)\agt10^{-2}$ and $10^{-3}$, for successful detection are much more strict than those in the actual experiments for electronic systems~\cite{Hafez2018,Cheng2020,Kovalev2020}.
Thus, much weaker pulses might be enough to verify our predictions experimentally.
These estimates indicate that lower-order harmonics in Kitaev magnets can be detected with current THz light techniques.

\textit{Conclusions.}-- We have analyzed the HHG in Kitaev magnets with an ME coupling
by using a quantum master equation and the linear response theory.
Our results show that the specific nature of the Majorana fermion excitations
can be detected by linear and nonlinear THz-light responses.
The characteristics of the Kitaev model, such as even-order harmonics, continuum HHG spectra, and broad peaks,
can be controlled by applying static electric or magnetic fields.
This study sheds light on an interdisciplinary field between photo science and QSLs.

Our setup does not accompany vison (localized fermion) excitations.
Studies for laser-driven vison dynamics and the effects of temperature and ac Zeeman terms
are interesting directions for future work.

\begin{acknowledgments}
	M. S. was supported by JSPS KAKENHI (Grant No. 17K05513 and No. 20H01830) and
	a Grant-in-Aid for Scientific Research on Innovative Areas
	``Quantum Liquid Crystals'' (Grant No. JP19H05825).
	T.N.I. was supported by JSPS KAKENHI Grant No. JP18K13495 and No. 21K13852.
\end{acknowledgments}



%


\setcounter{figure}{0}
\setcounter{equation}{0}
\setcounter{section}{0}

\onecolumngrid

\begin{center}

	\vspace{1.5cm}

	{\large \textbf{Supplemental Material: Linear and Nonlinear Optical Responses in Kitaev Spin Liquids}}

	\vspace{0.3cm}

	{\large Minoru~Kanega,${}^1$~Tatsuhiko~N.~Ikeda,${}^2$~Masahiro~Sato$^1$} \\[2mm]
	\textit{${}^1$Department of Physics, Ibaraki University, Mito,
		Ibaraki 310-8512, Japan}\\
	\textit{${}^2$Institute for Solid State Physics, University of Tokyo, Kashiwa, Chiba 277-8581, Japan}\\

\end{center}

\vspace{0.6cm}

\renewcommand{\thesection}{S\arabic{section}}
\renewcommand{\theequation}{S\arabic{section}.\arabic{equation}}
\renewcommand{\thefigure}{S\arabic{figure}}
\renewcommand{\thetable}{S\arabic{table}}

\makeatletter
\@addtoreset{equation}{section}
\makeatother

\section{S1. FERMIONIZATION of KITAEV HONEYCOMB MODEL}
\setcounter{section}{1}

\begin{figure}[htb]
  \centering
  \includegraphics[width=0.7\linewidth]{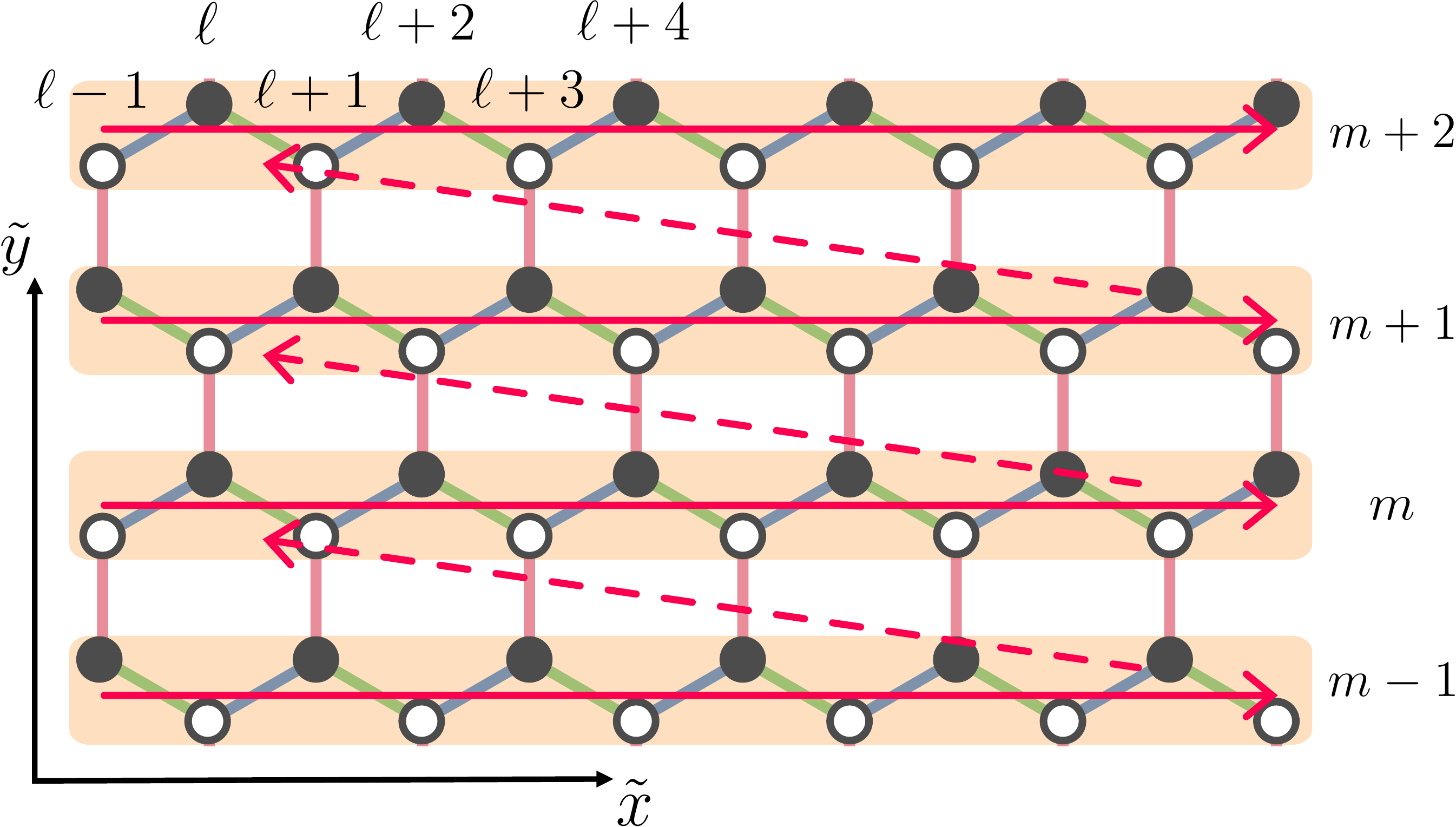}
  \caption{Honeycomb lattice structure of the Kitaev model of Eq.\autoref{eq:hamiltonian}.
    Jordan--Wigner (JW) transformation is performed along the red lines.}
  \label{fig:JWtrans}
\end{figure}

In this section, we shortly review the fermionization of the Kitaev model~\cite{Kitaev2006a,Trebst2017,Hermanns2018,Takagi2019,Motome2020a}.
The Hamiltonian of the Kitaev model is given by
\begin{align}
  H_0 & =  H_K + \hms + H_{\kappa} \notag                                                              \\
      & =-J\sum_{\alpha=\{x,y,z\}}\sum_{\braket{\br,\br'}_\alpha}\sigma^\alpha_\br\sigma^\alpha_{\br'}
  -\edc\qty(-\sum_{\braket{\br,\br'}_x}\sigma^x_\br\sigma^x_{\br'}+\sum_{\braket{\br,\br'}_y}\sigma^y_\br\sigma^y_{\br'})
  -\kappa\sum_{\mathrm{NNN}}\sigma^x_\br\sigma^y_{\br'}\sigma^z_{\br''},
  \label{eq:hamiltonian}
\end{align}
where $\sigma^\alpha_\br$ is the $\alpha$ component of Pauli matrix on site $\br$.
The first term $H_K$ is the Kitaev exchange interaction on nearest-neighboring bonds.
The second one $\hms=E_\dc\eta_{\mathrm{ms}}(-\sum_{\braket{\br,\br'}_x}\sigma^x_\br\sigma^x_{\br'}+\sum_{\braket{\br,\br'}_y}\sigma^y_\br\sigma^y_{\br'})$ with $\edc=E_\dc\eta_{\mathrm{ms}}$ is the dimerization energy
driven by a static electric field $E_\dc$ along the $\tilde x$ direction (see Fig.~\hyperref[fig:band]{1} in the main text).
The dimerization is proportional to the electric polarization $\hat{P}$ as
\begin{align}
  \hat{P} & = \eta_{\mathrm{ms}} \qty(-\sum_{\braket{\br,\br'}_x}\sigma^x_\br\sigma^x_{\br'}+\sum_{\braket{\br,\br'}_y}\sigma^y_\br\sigma^y_{\br'}),
\end{align}
where $\eta_{\mathrm{ms}}$ is the magnetoelectric (ME) coupling constant~\cite{Tokura2014}.
The final term $H_{\kappa}$ is the three-spin term~\cite{Kitaev2006a} driven by a static magnetic field ${\bm{B}}$,
and we will explain $H_{\kappa}$ in more detail later.
This model can be fermionized by a Jordan--Wigner (JW) transformation,
by regarding it as the array of one-dimensional chains consisting of the $x$ and $y$ bonds~\cite{Chen2008}.
See Fig.\autoref{fig:JWtrans}.
Through the JW transformation, the spin operators are rewritten by using spinless fermions as
\begin{equation}
  \left\{
  \begin{aligned}
    \sigma^+_{jk} & =c_{jk}^{\dag}\exp\qty[i\pi\qty(\sum_{m=1}^{k-1}\sum_{\ell}c_{\ell m}^{\dag}c_{\ell m}+\sum_{\ell =1}^{j-1}c_{\ell k}^{\dag}c_{\ell k})], \\
    \sigma^-_{jk} & =c_{jk}\exp\qty[i\pi\qty(\sum_{m=1}^{k-1}\sum_{\ell}c_{\ell m}^{\dag}c_{\ell m}+\sum_{\ell=1}^{j-1}c_{\ell k}^{\dag}c_{\ell k})],         \\
    \sigma^z_{jk} & =2c_{jk}^{\dag}c_{jk}-1,
  \end{aligned}
  \right.
  \label{eq:2dimJW}
\end{equation}
where $\sigma^{\pm}_\br=(\sigma^{x}_\br \pm i \sigma^{y}_\br)/2$.
$c^\dag_\br$ and $c_\br$ are respectively the spinless-fermion creation and annihilation operators
that satisfy the anticommutation relations,
\begin{equation}
  \acomm{c^\dag_\br}{c_{\br'}}=\delta_{\br\br'},\quad\acomm{c^\dag_\br}{c^\dag_{\br'}}=0,\quad\acomm{c_\br}{c_{\br'}}=0.
\end{equation}

\begin{figure}[htb]
  \centering
  \includegraphics[width=0.7\linewidth]{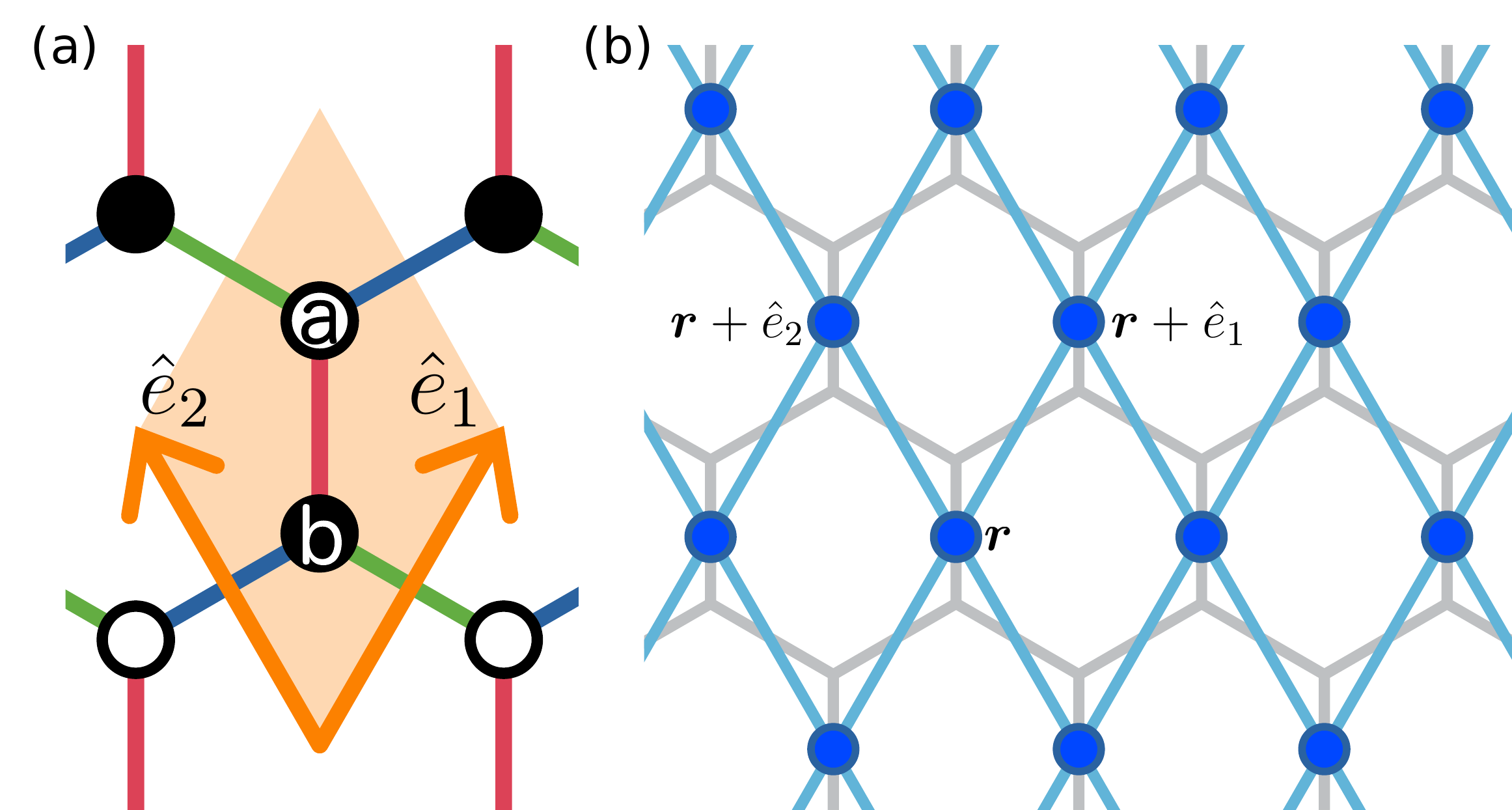}
  \caption{(a) Unitcell of the Kitaev model. $\hat{e}_1$ and $\hat{e}_2$ are primitive translation vectors.
    (b) Lattice structure for the complex fermions $d_\br$ of the Kitaev model.
    The fermion $d_\br$ is defined on the virtual site (blue site) which is the bond center of each $z$ bond.}
  \label{fig:complexFermionization}
\end{figure}

First, we apply the JW transformation\autoref{eq:2dimJW} to $H_K+\hms$:
\begin{align}
  H_K+\hms  = & -(J-\edc)\sum_{\braket{\br,\br'}_x}\qty((c\brw)^\dag-c\brw)\qty((c\brbp)^\dag+c\brbp)+(J+\edc)\sum_{\braket{\br,\br'}_y}\qty((c\brb)^\dag+c\brb)\qty((c\brwp)^\dag-c\brwp) \notag \\
              & -J \sum_{\braket{\br,\br'}_z}\qty((c\brb)^\dag-c\brb)\qty((c\brb)^\dag+c\brb)\qty((c\brwp)^\dag-c\brwp)\qty((c\brwp)^\dag+c\brwp).
  \label{eq:fermionization}
\end{align}
In Eq.\autoref{eq:fermionization}, the symbol $\br$ is re-defined as a vector pointing to the bond center of $z$ bond,
and the index $a$ ($b$) denotes the sublattice $a$ ($b$) on the same $z$ bond,
as shown in Fig.\autoref{fig:complexFermionization}(a).
Here, we introduce the Majorana (real) fermion from the linear combination of $c_\br^\dag$ and $c_\br$:
\begin{align}
   & \xi\brw=\frac{\qty(c\brw-(c\brw)^\dag)}{i},\quad \chi\brw=\qty(c\brw+(c\brw)^\dag), \nonumber \\
   & \chi\brb=\frac{\qty(c\brb-(c\brb)^\dag)}{i},\quad \xi\brb=\qty(c\brb+(c\brb)^\dag).
  \label{eq:majorana}
\end{align}
Substituting Eq.\autoref{eq:majorana} into Eq.\autoref{eq:fermionization},
we can rewrite Hamiltonian\autoref{eq:fermionization} as
\begin{equation}
  H_K+\hms=i(J-\edc)\sum_{\braket{\br,\br'}_x}\xi\brw \xi\brbp-i(J+\edc)\sum_{\braket{\br,\br'}_y}\xi\brb \xi\brwp-iJ\sum_{\braket{\br,\br'}_z}\alpha_{\br}\xi\brb \xi\brwp,
  \label{eq:majoranaHamiltonian}
\end{equation}
where $\alpha_\br=i\chi\brb \chi\brwp$ with $\br$ and $\br'$ being on a $z$ bond.
The operator $\alpha_\br$ in each $z$ bond is the local conserved quantity
and has eigenvalues $1$ or $-1$. In the ground state, the spatial distribution of the eigenvalue of $\{\alpha_\br\}$
is shown to be all unity (or all -1)~\cite{Lieb1994}.
Therefore, the pair $(\xi\brb,\xi\brwp)$ correspond to itinerant (dispersive) fermions,
while the remaining fermions $(\chi\brb, \chi\brwp)$ are localized excitations (called vison).
The vison has no dynamics and its excitation gap is estimated as $\sim0.07|J|$ in the isotropic Kitaev model
with $J_{x,y,z}=J$~\cite{Kitaev2006a}.
From these arguments, the ground-state Hamiltonian (i.e., zero temperature $T=0$)
is described by only the itinerant Majorana fermions.
We note that if we perform the perturbation theory for some of realistic perturbations, such as Heisenberg-term
\begin{equation}
    H_{J_H}=J_H\sum_{\braket{\br,\br'}}\bm{\sigma}_\br\cdot\bm{\sigma}_{\br'},
\end{equation}
$\Gamma$-term
\begin{equation}
    H_\Gamma=-\Gamma\sum_{\substack{\alpha,\beta,\gamma=\{x,y,z\}\\\beta\neq\alpha,\gamma\neq\alpha}}\sum_{\braket{\br,\br'}_\alpha}\qty[\sigma^\beta_\br\sigma^\gamma_{\br'}+\sigma^\gamma_\br\sigma^\beta_{\br'}],
\end{equation}
and $\Gamma'$-term
\begin{equation}
    H_{\Gamma'}=-\Gamma'\sum_{\substack{\alpha,\beta=\{x,y,z\}\\\beta\neq\alpha}}\sum_{\braket{\br,\br'}_\alpha}\qty[\sigma^\alpha_\br\sigma^\beta_{\br'}+\sigma^\beta_\br\sigma^\alpha_{\br'}],
\end{equation}
their leading terms can be mapped to
itinerant Majorana fermions~\cite{Takikawa2019,Takikawa2020}.
Namely, such perturbed Kitaev models can also be analyzed by using the fermionization. We note that the three-spin $\kappa$ term is also derived from the third-order perturbation calculation in terms of a standard Zeeman interaction~\cite{Kitaev2006a}.

In the present study, we can choose $\alpha_\br=1$ without loss of generality,
and focus on the ground state below.
From the itinerant Majorana fermions, we define a complex fermion on the new site $\br$ which is center of each $z$ bond
  [see Fig.\autoref{fig:complexFermionization}(b)]:
\begin{equation}
  \left\{
  \begin{aligned}
     & d_\br        =\frac{1}{2}\qty(\xi\brb+i\xi\brw), \\
     & d^\dag_\br  =\frac{1}{2}\qty(\xi\brb-i\xi\brw),
  \end{aligned}
  \right.
  \label{eq:reducedFermion}
\end{equation}
where $d^\dag_\br$ and $d_\br$ are respectively the spinless-fermion creation and annihilation operators
that satisfy the anticommutation relations,
\begin{equation}
  \acomm{d^\dag_\br}{d_{\br'}}=\delta_{\br\br'},\quad\acomm{d^\dag_\br}{d^\dag_{\br'}}=0,\quad\acomm{d_\br}{d_{\br'}}=0.
\end{equation}

Substituting Eq.\autoref{eq:reducedFermion} into Eq.\autoref{eq:majoranaHamiltonian}, we represent
the ground-state Hamiltonian with $\{\alpha_\br\}=1$ as
\begin{equation}
  H_K+\hms=-\sum_\br\qty[(J-\edc)\qty(d_\br^\dag-d_\br)\qty(d^\dag_\brx+d_\brx)+(J+\edc)\qty(d_\br^\dag-d_\br)\qty(d^\dag_\bry+d_\bry)+J\qty(2 d_\br^\dag d_\br-1)],
  \label{eq:GShamiltonian}
\end{equation}
where $\hat{e}_{1}$ and $\hat{e}_{2}$ are primitive translation vectors of the Kitaev model, as shown
in Fig.\autoref{fig:complexFermionization}.
Through Fourier transformation for the complex fermion $d_\br$,
\begin{equation}
  d_\br=\frac{1}{\sqrt{N}}\sum_\bk\tilde{d}_\bk e^{i\bk\cdot\br},
\end{equation}
we obtain
\begin{equation}
  H_K+\hms=\sum_{\substack{\bk \\ k_x>0}}\qty[\epsilon_\bk \tild_\bk^\dag \tild_\bk-\epsilon_\bk \tild_{-\bk} \tild_{-\bk}^\dag-i\Delta^0_\bk\qty(\tild_\bk^\dag \tild_{-\bk}^\dag-\tild_{-\bk}\tild_\bk)],
  \label{eq:hk+Hms}
\end{equation}
where $\bk=(k_x,k_y)^\top$ is a wave vector.
$k_x=2\pi(n/N_1-m/N_2)/\sqrt{3}$ and $k_y=2\pi(n/N_1+m/N_2)/3\;\;(n,m\in\mathbb{Z})$
are respectively the wave numbers along the $\tilde{x}$ and $\tilde{y}$ directions,
where $N=N_1N_2$ is the total number of unit cells and $N_{1(2)}$ is the size along the $\hat{e}_{1(2)}$ direction.
The parameters $\epsilon_\bk$ and $\Delta^0_\bk$ are respectively defined as
$\epsilon_\bk=\epsilon^{\mathrm{iso}}_\bk-E_\dc\epsilon^P_\bk$ and
$\Delta^0_\bk=\Delta^{\mathrm{iso}}_\bk-E_\dc\Delta^P_\bk$,
in which
\begin{align}
   & \epsilon^{\mathrm{iso}}_\bk=-2J\qty(1+\cos(\bk\cdot\hat{e}_1)+\cos(\bk\cdot\hat{e}_2)), \\
   & \Delta^{\mathrm{iso}}_\bk=2J\qty(\sin(\bk\cdot\hat{e}_1)+\sin(\bk\cdot\hat{e}_2)),      \\
   & \epsilon^P_\bk=-2\etams\qty(\cos(\bk\cdot\hat{e}_1)-\cos(\bk\cdot\hat{e}_2)),           \\
   & \Delta^P_\bk=2\etams\qty(\sin(\bk\cdot\hat{e}_1)-\sin(\bk\cdot\hat{e}_2)).
\end{align}
For simplicity, hereafter we will rewrite $\sum_{\bk,k_x>0}$ as $\sum'_\bk$.

\begin{figure}[htb]
  \centering
  \includegraphics[width=0.6\linewidth]{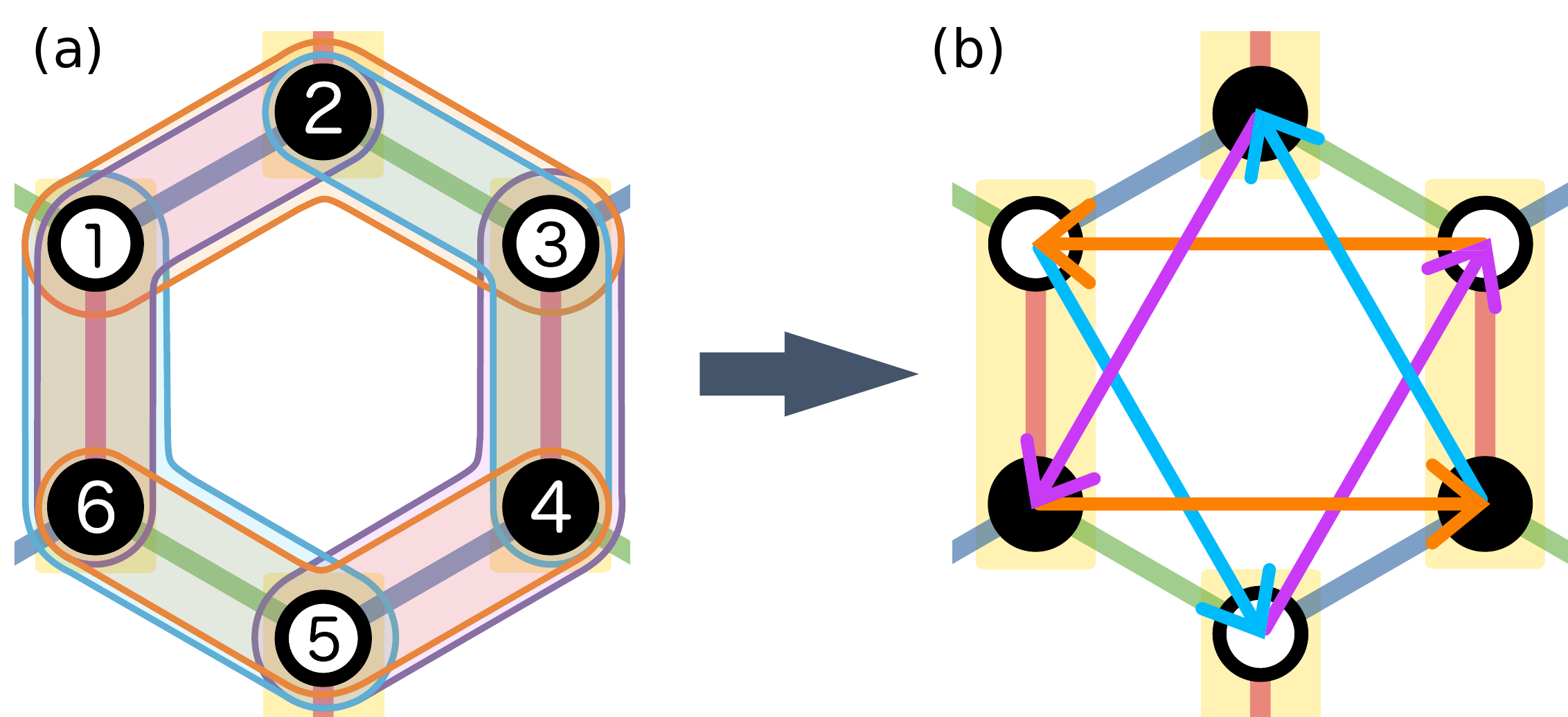}
  \caption{(a) Six kinds of three-spin interactions in each plaquette. Three-spin interactions are defined on
    trios of three neighboring sites: $\{1,2,3\}$,  $\{2,3,4\}$, $\{3,4,5\}$, $\{4,5,6\}$, $\{5,6,1\}$, and $\{6,1,2\}$.
    (b) Fermionized $\kappa$ term in a plaquette.
    Each three-spin term is mapped to a next-nearest neighboring hopping with an imaginary factor.
    Namely, the three-spin term provides an effective ``magnetic flux'' for spinless fermions.}
  \label{fig:kappa-config}
\end{figure}

Next let us consider the three-spin interaction $H_{\kappa}$. Each honeycomb plaquette has
six kinds of three-spin terms defined on six trios consisting of three neighboring sites:
$\{1,2,3\}$,  $\{2,3,4\}$, $\{3,4,5\}$, $\{4,5,6\}$, $\{5,6,1\}$, and $\{6,1,2\}$.
These six triplets are shown in Fig.\autoref{fig:kappa-config}(a).
Following the fashion respecting the honeycomb plaquettes, we represent the three-spin interaction $H_{\kappa}$ as
\begin{align}
  H_{\kappa}= & -\kappa\sum_\bp\left(\sigma^x_{\bp,1a}\sigma^z_{\bp,2b}\sigma^y_{\bp,3a}+\sigma^z_{\bp,6b}\sigma^y_{\bp,1a}\sigma^x_{\bp,2b}+\sigma^y_{\bp,2b}\sigma^x_{\bp,3a}\sigma^z_{\bp,4b}\right.\notag \\
              & \qquad\qquad\left.+\sigma^x_{\bp,4b}\sigma^z_{\bp,5a}\sigma^y_{\bp,6b}+\sigma^x_{\bp,5a}\sigma^y_{\bp,4b}\sigma^z_{\bp,3a}+\sigma^z_{\bp,1a}\sigma^x_{\bp,6b}\sigma^y_{\bp,5a}\right),
\end{align}
where $\sum_\bp$ means the sum over all the plaquettes. Through the JW transformation, it can be mapped to
the next-nearest-neighboring (NNN) hopping terms of Majorana fermions (see Fig.\autoref{fig:kappa-config}(b)),
\begin{align}
  H_{\kappa}= & -\kappa\sum_\bp i\left(\xi_{\bp}^{6b}\xi_{\bp}^{4b}+\alpha_{\bp,6-1}\xi_{\bp}^{2b}\xi_{\bp}^{6b}+\alpha_{\bp,4-3}\xi_{\bp}^{4b}\xi_{\bp}^{2b}\right.\notag \\
              & \qquad\qquad\left.+\xi_{\bp}^{3a}\xi_{\bp}^{1a}+\alpha_{\bp,4-3}\xi_{\bp}^{5a}\xi_{\bp}^{3a}+\alpha_{\bp,6-1}\xi_{\bp}^{1a}\xi_{\bp}^{5a}\right),
\end{align}
where $\alpha_{\bp,4-3}=i\chi_\bp^{4b} \chi_\bp^{3a}$ and $\alpha_{\bp,6-1}=i\chi_\bp^{6b} \chi_\bp^{1a}$.
In the ground state, its complex-fermion representation is given by
\begin{align}
  H_{\kappa}= & -i\kappa\sum_\br \left[\qty(d_\bry+d^\dag_\bry)\qty(d_\brx+d^\dag_\brx)+\qty(d_\brx+d^\dag_\brx)\qty(d_\br+d^\dag_\br)+\qty(d_\br+d^\dag_\br)\qty(d_\bry+d^\dag_\bry)\right.\notag \\
              & \left.\qquad\qquad-\qty(d_\brym-d^\dag_\brym)\qty(d_\brxm-d^\dag_\brxm)-\qty(d_\brxm-d^\dag_\brxm)\qty(d_\br-d^\dag_\br)-\qty(d_\br-d^\dag_\br)\qty(d_\brym-d^\dag_\brym)\right],
  \label{eq:kappaHamiltonian}
\end{align}
where we have used $\{\alpha_\br\}=1$.
Since this NNN hopping have an imaginary factor $\propto i\kappa$, it plays the role of an effective ``magnetic flux''
for the fermion $\{d_{\br}\}$.
Applying Fourier transformation to Eq.\autoref{eq:kappaHamiltonian}, we obtain
\begin{equation}
  H_{\kappa}=\sumk\Delta^B_\bk\qty(\tild^\dag_\bk \tild^\dag_{-\bk}+\tild_{-\bk}\tild_\bk),
  \label{eq:Hkappa}
\end{equation}
where
\begin{equation}
  \Delta^B_\bk=4\kappa\qty[\sin(\sqrt{3}k_x)-\sin(\bk\cdot\hat{e}_1)+\sin(\bk\cdot\hat{e}_2)].
\end{equation}

From Eqs.\autoref{eq:hk+Hms} and\autoref{eq:Hkappa},
the Hamiltonian $H_0 =  H_K + \hms + H_{\kappa}$ is expressed in the following matrix form,
\begin{equation}
  H_0=\sumk\bm{D}_\bk^\dag M_\bk\bm{D}_\bk,
  \label{eq:totalkHam}
\end{equation}
where
\begin{equation}
  M_\bk=\mqty(\epsilon_\bk    & \Delta_\bk \\ \Delta_\bk^* & -\epsilon_\bk),
  \qquad\bm{D}_\bk=\mqty(\tilde{d}_\bk & \tilde{d}^\dag_{-\bk})^\top,
\end{equation}
with $\Delta_\bk=-i\Delta^0_\bk+\Delta^B_\bk$.
This is the same form as that of a BCS Hamiltonian for superconductors.
Through unitary transformation $\mqty(f_\bk & f^\dag_{-\bk})^\top=U^\dag_\bk \bm{D}_\bk$,
Eq.\autoref{eq:totalkHam} is diagonalized as
\begin{equation}
  H_0=\sumk E_\bk(f^\dag_\bk f_\bk+f_{-\bk}f^\dag_{-\bk}),
\end{equation}
with the energy band
\begin{equation}
  E_\bk=\sqrt{\epsilon_\bk^2+\abs{\Delta_\bk}^2}.
\end{equation}
We have introduced new fermions $\{f_\bk\}$ and unitary matrices $U_\bk$.
The ground state is given by
\begin{equation}
  \ket{\mathrm{gs}}=\sideset{}{'}\prod_{\bk}(u_\bk-v_\bk\tilde{d}^\dag_\bk\tilde{d}^\dag_{-\bk})\ket{0},
\end{equation}
with
\begin{align}
  u_\bk =\sqrt{\frac{1}{2}\qty(1+\frac{\epsilon_\bk}{E_\bk})}, & \qquad
  v_\bk =\frac{\Delta_\bk}{\qty|\Delta_\bk|}\sqrt{\frac{1}{2}\qty(1-\frac{\epsilon_\bk}{E_\bk})},
\end{align}
where $\prod'_{\bk}=\prod_{\bk,k_x>0}$, and $\ket{0}$ is the Fock vacuum for the fermion $\{\tilde{d}_\bk\}$.

The electric polarization $\hat{P}$ can also be fermionized through JW transformation since it is a part of the Hamiltonian.
The result is as follows:
\begin{equation}
  \hat{P}=\sumk\bm{D}_\bk^\dag P_\bk\bm{D}_\bk,
\end{equation}
with
\begin{equation}
  P_\bk=\mqty(\epsilon^P_\bk    & -i\Delta^P_\bk \\ i\Delta^P_\bk & -\epsilon^P_\bk).
\end{equation}

\section{S2. GROUND STATE ENERGY and DC ELECTRIC FIELD DEPENDENCE}
\setcounter{section}{2}
\setcounter{equation}{0}

\begin{figure}[htb]
  \centering
  \includegraphics[width=0.6\linewidth]{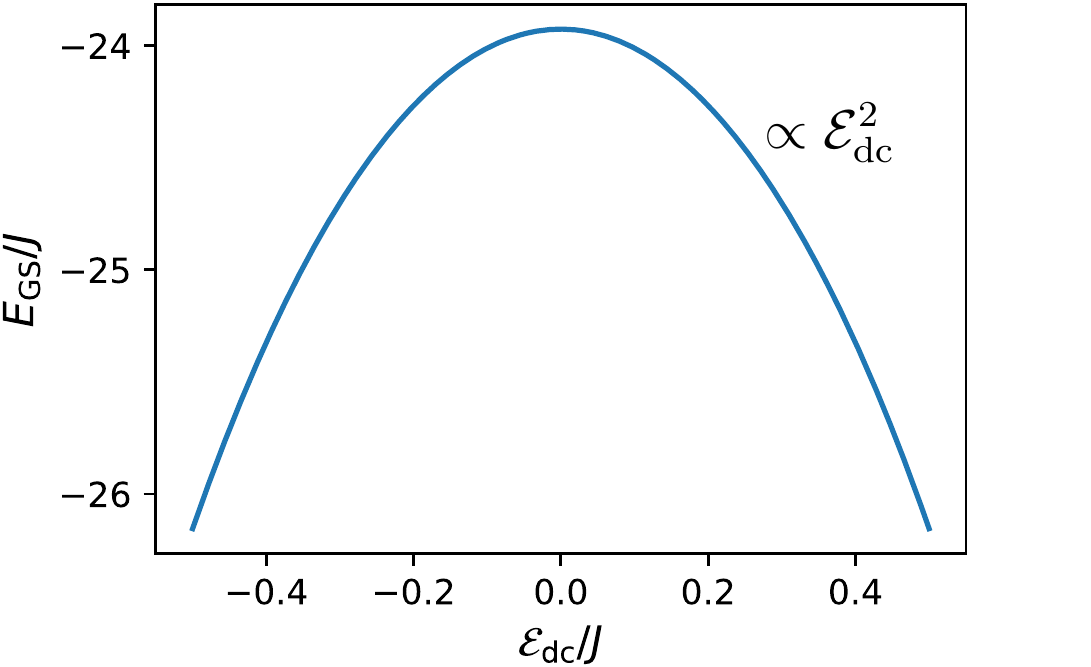}
  \caption{$\edc$ dependence of the ground-state energy $E_{\mathrm{GS}}$
    in the Kitaev model with $\kappa=0$.}
  \label{fig:GS-energy}
\end{figure}

From the above section, the Hamiltonian of the Kitaev model is fermionized as
\begin{equation}
  H_0=\sumk E_\bk(f^\dag_\bk f_\bk+f^\dag_{-\bk}f_{-\bk})-\sumk E_\bk.
\end{equation}
Therefore, the ground-state energy per one unit cell is given by $E_{\mathrm{GS}}=-N^{-1}\sum'_\bk E_\bk$
with $N$ being the total number of unit cells. From the numerical calculation of $E_{\mathrm{GS}}$,
we plot the ground-state energy as the function of dc electric field $E_\dc$ in Fig.\autoref{fig:GS-energy}.
It shows that if a dc electric field is applied, the energy decreases with being proportional to $\edc^2$.
This means that when the system is distorted by a dc electric field $E_\dc$, the electric state becomes more stable.
This result is consistent with our assumption that the dimerization is coupled to an applied electric field,
and it indicates that the ME coupling can appear in a class of Kitaev-magnet candidates.

\section{S3. TIME EVOLUTION of DENSITY MATRIX via LINDBLAD EQUATION}
\setcounter{section}{3}
\setcounter{equation}{0}

In this section, we explain the quantum master equation approach~\cite{Breuer2007,Alicki2007,Ikeda2019,Sato2020} to our driven Kitaev model.
We consider a Kitaev model under the ac ME coupling between the polarization
and an applied ac electric field $E_\ac(t)$ along the $\tilde x$ direction.
The time-dependent Hamiltonian is given by
\begin{equation}
  H(t) = H_0-E_\ac(t)\hat{P}.
  \label{eq:totalHamiltonian}
\end{equation}
Hereafter, we choose two vectors
$\tilde{d}^\dag_\bk\tilde{d}^\dag_{-\bk}\ket{0_\bk}$ and $\ket{0_\bk}$ as the bases in each subspace with $\bk$.
Here, $\ket{0_\bk}$ is the vacuum of the fermion $\tilde{d}_\bk$.
Using these bases, we can express the Hamiltonian\autoref{eq:totalHamiltonian} in the $2\times 2$ form,
\begin{align}
  H(t)                          & =\sumk\qty[\mqty(\epsilon_\bk & \Delta_\bk \\ \Delta_\bk^* & -\epsilon_\bk)
  -E_\ac(t)\mqty(\epsilon^P_\bk & -i\Delta^P_\bk                             \\ i\Delta^P_\bk & -\epsilon^P_\bk)]
  =\sumk\qty(M_\bk-E_\ac(t)P_\bk).
\end{align}
We note that in our setup, there is no vison dynamics even when the laser is applied.
Namely, the Hamiltonian for the laser-driven dynamics includes only itinerant Majorana fermions.

To describe the realistic non-equilibrium dynamics for the driven Kitaev model,
we use the following quantum master equation of the Lindblad form~\cite{Breuer2007,Alicki2007,Ikeda2019,Sato2020}
\begin{equation}
  \frac{d\rho(\bk,t)}{dt}=-i\qty[h(\bk,t),\rho(\bk,t)]+\gamma\qty(L_\bk\rho(\bk,t)L_\bk^{\dag}-\frac{1}{2}\qty{L_\bk^{\dag}L_\bk,\rho(\bk,t)}),
  \label{eq:lindblad}
\end{equation}
where $\rho(\bk,t)$ is the density matrix in the $\bk$ subspace.
$h(\bk,t)=M_\bk-E_\ac(t)P_\bk$ is the Hamiltonian in the $\bk$ subspace and
the first commutator of Eq.\autoref{eq:lindblad} represents the unitary dynamics caused by the Hamiltonian.
The second $\gamma$ term describes the dissipation dynamics, and it is necessary
to describe ac-field driven phenomena in materials because dissipation effect is usually inevitable in most of matter.
We set Lindblad operator $L_\bk$ to be
\begin{equation}
  L_\bk=\ket{g_\bk}\bra{e_\bk},
\end{equation}
where $\ket{g_\bk}=(-v_\bk,u_\bk)^\top$ and $\ket{e_\bk}=(u_\bk,v^*_\bk)^\top$ are respectively
ground and excited states of the static part of the Hamiltonian in the subspace $\bk$.
The Lindblad operator describes the relaxation to the ground state, and $1/\gamma$ is viewed as the relaxation time.
For simplicity, we have assumed that the relaxation processes in multiple subspaces with different $\bk$
are independent of each other, and we have used a single value of $\gamma$ for the total system.
Unless otherwise stated throughout this work, we use a typical value $\gamma=0.1J$, which corresponds to the relaxation time $\tau=1/\gamma\sim\SI{7.6}{ps}$ for $J/k_B=\SI{10}{K}$. We will shortly discuss the relaxation-rate dependence
of the HHG spectra in Sec.~\hyperref[sec:relaxation]{S4}.
The initial density matrix $\rho(\bk,t_{\mathrm{ini}})$ at time $t=t_{\mathrm{ini}}$ is set to be
\begin{equation}
  \rho(\bk,t_{\mathrm{ini}})=\ket{g_\bk}\bra{g_\bk}.
\end{equation}

To simplify the description,
we perform a unitary transformation to the Lindblad equation\autoref{eq:lindblad}
with the unitary matrix
\begin{equation}
  U_\bk=\mqty(\ket{e_\bk} & \ket{g_\bk})=\mqty(u_\bk & -v_\bk \\ v^*_\bk & u_\bk),
\end{equation}
which makes the matrix $M_\bk$ diagonalized.
Using $U_\bk$, we define some matrices as follows:
\begin{equation}
  \left\{
  \begin{aligned}
     & U_\bk^\dag\rho(\bk,t)U_\bk=\tilde{\rho}(\bk,t),               \\
     & U_\bk^\dag M_\bk U_\bk=\mqty(E_\bk              & 0           \\ 0 & -E_\bk)=\tilde{M}_\bk, \\
     & U_\bk^\dag  P_\bk U_\bk=\mqty(p^{11}_\bk        & p^{12} _\bk \\ \qty(p^{12}_\bk)^* & -p^{11}_\bk)=\tilde{P}_\bk, \\
     & U_\bk^\dag L_\bk U_\bk=\mqty(0                  & 0           \\ 1 & 0)= \tilde{L}_\bk,
  \end{aligned}
  \right.
\end{equation}
where
\begin{equation}
  \left\{
  \begin{aligned}
     & p^{11}_\bk=\frac{\epsilon_\bk \epsilon^P_\bk+\Delta^0_\bk\Delta^P_\bk}{E_\bk},                                                                                                        \\
     & p^{12}_\bk=-\frac{\epsilon_\bk^P\Delta_\bk+i\Delta_\bk^P\epsilon_\bk}{E_\bk}-i\Delta^P_\bk\frac{\Delta^B_\bk\Delta_\bk}{\abs{\Delta_\bk}^2}\left(1-\frac{\epsilon_\bk}{E_\bk}\right). \\
  \end{aligned}
  \right.
\end{equation}
In addition, using properties of Pauli matrices $\sigma_{x,y,z}$ (note that they are not spin operators),
we can rewrite $\tilde{\rho}(\bk,t)$, $\tilde{h}(\bk,t)=\tilde{M}_\bk-E_\ac(t)\tilde{P}_\bk$, and $\tilde{P}_\bk$
as follows:
\begin{equation}
  \left\{
  \begin{aligned}
     & \tilde{\rho}(\bk,t)=\frac{1+\sum_{\alpha}\tau^\alpha_\bk(t)\sigma_\alpha}{2}, \\
     & \tilde{h}(\bk,t)(t)=\sum_\alpha h^\alpha_\bk(t)\sigma_\alpha,                 \\
     & \tilde{P}_\bk(t)=\sum_\alpha p^\alpha_\bk\sigma_\alpha,
  \end{aligned}
  \right.
\end{equation}
where $\tau^\alpha_\bk(t)$, $h^\alpha_\bk(t)$, and $p^\alpha_\bk$ ($\alpha=x,y,z$) are given by
\begin{equation}
  \left\{
  \begin{aligned}
     & \tau^\alpha_\bk(t)=\Tr(\tilde{\rho}(\bk,t)\sigma_\alpha),        \\
     & h^\alpha_\bk(t)=\frac{\Tr(\tilde{h}(\bk,t)(t)\sigma_\alpha)}{2}, \\
     & p^\alpha_\bk=\frac{\Tr(\tilde{P}_\bk\sigma_\alpha)}{2}.
  \end{aligned}
  \right.
\end{equation}
Through these instruments, the Lindblad equation\autoref{eq:lindblad} for the density matrix $\rho(\bk,t)$
is rewritten in the following three-component vector form,
\begin{equation}
  \frac{d\bm{\tau}_\bk(t)}{dt}=2\bm{h}_\bk(t)\times\bm{\tau}_\bk(t)+\gamma\qty(-\bm{\tau}_\bk(t)+\frac{1}{2}\mqty(\tau_\bk^x(t)\\\tau_\bk^y(t)\\-2)).
\end{equation}
The initial state is given by
\begin{equation}
  \bm{\tau}_\bk(t_{\mathrm{ini}})=\mqty(0\\0\\-1).
\end{equation}
With $\tau^\alpha_\bk(t)$, the expectation value of the polarization is also rewritten as
\begin{align}
  P(t)=\braket{\hat{P}}_t & =\frac{1}{N}\sumk\Tr\qty(\rho(\bk,t)P_\bk)\notag            \\
                          & =\frac{1}{N}\sumk\sum_\alpha\tau^\alpha_\bk(t)p^\alpha_\bk.
\end{align}
In our numerical calculation, the total number of unit cells, $N$, is set to be $N=160\times160$.

\section{S4. RELAXATION-RATE DEPENDENCE of HHG SPECTRA}
\setcounter{section}{4}
\setcounter{equation}{0}
\label{sec:relaxation}

\begin{figure}[htbp]
    \centering
    \includegraphics[width=\linewidth]{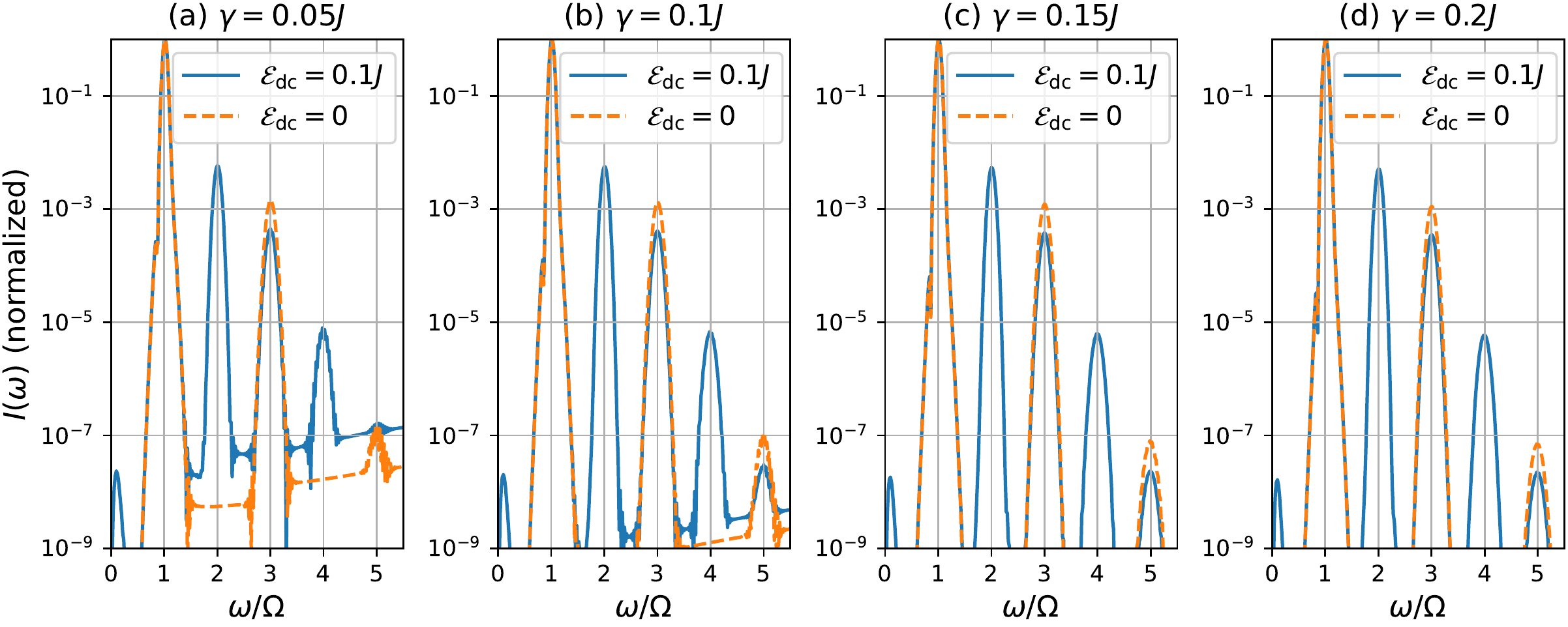}
    \caption{
    HHG spectra $I(\omega)$ in the driven ferromagnetic ($J>0$) Kitaev models under a THz pulse with
    $\Omega=2.0J$ and $\eac=0.1J$ as a function of $\omega$ at $\edc=0$ and $0.1J$.
    Panels (a)--(d) correspond to the results of $\gamma/J=(0.05, 0.1, 0.15, 0.2)$, respectively.
    $I(\omega)$ is normalized with its maximum value.}
    \label{fig:relaxation}
\end{figure}

As we mentioned in the above section, here we discuss the relaxation-rate dependence of the HHG spectra
around the typical value $\gamma=0.1J$.
In Fig.\autoref{fig:relaxation}, we depicted the numerical results of the spectra under different relaxation rates
$\gamma/J=0.05$, $0.1$, $0.15$, and $0.2$.
Figure\autoref{fig:relaxation} shows that smaller $\gamma$ (longer relaxation time) causes a larger background noise,
while the peak positions and intensities of the HHG spectra are robust against moderately change of the
value of $\gamma$ around $\gamma=0.1J$. Therefore, we can expect that
our numerical results with $\gamma=0.1J$ are basically stable against a small change of the relaxation rate.

\section{S5. LINEAR RESPONSE THEORY for POLARIZATION}
\setcounter{section}{5}
\setcounter{equation}{0}

In this section, we explain the linear response of the polarization $\hat{P}$
to an ac electric field $E_\ac(t)$.
According to linear response theory, the Fourier transform of the linear response of $\hat{P}$
is given by $P(\omega)=\chi_P(\omega)E_\ac(\omega)$
when the external field $E_\ac(\omega)$ with frequency $\omega$ is sufficiently weak.
The susceptibility $\chi_P(\omega)$ is calculated from the imaginary-time
two-point function of the polarization with analytic continuation
$i\omega_n\rightarrow\omega+i\delta$ ($\omega_n$ is Matsubara frequency).
Below we show the calculation of $\chi_P(\omega)$.
Let us first define the imaginary-time two-point function at finite temperature as
\begin{equation}
  \Phi_0(\tau,\tau')=\frac{1}{N}\Braket{\ttau{\hat{P}(\tau)\hat{P}(\tau')}},
  \label{eq:imTimeCorr}
\end{equation}
where $\tau$ is the imaginary time, $N$ is the total number of unit cells,
and $\ttau{\cdots}$ means the imaginary-time ordered product.
The $\tau$-dependent polarization $\hat{P}(\tau)$ is given by
\begin{equation}
  \hat{P}(\tau)=\sumk\hat{\bm{F}}^\dag_\bk(\tau)\tilde{P}_\bk\hat{\bm{F}}_\bk(\tau),
\end{equation}
where $\hat{\bm{F}}_\bk(\tau)=\mqty(\hat{f}_\bk(\tau) & \hat{f}^\dag_{-\bk}(\tau))^\top$ and
$\hat{f}^\dag_\bk(\tau)$ ($\hat{f}_\bk(\tau)$) is the fermion creation (annihilation) operator
in the imaginary-time Heisenberg representation.

Applying Bloch--De Dominicis theorem to Eq.\autoref{eq:imTimeCorr}, we obtain
\begin{equation}
  \Phi_0(\tau,\tau')=\frac{1}{N}\sumk\qty|p^{12}_\bk|^2\qty(\gz(\bk,\tau'-\tau)\gz(-\bk,\tau'-\tau)+\gz(\bk,\tau-\tau')\gz(-\bk,\tau-\tau')),
  \label{eq:suscepImTime}
\end{equation}
where $\gz(\bk,\tau)$ is the imaginary-time Green's function
\begin{equation}
  \gz(\bk,\tau-\tau')=-\Braket{\ttau{\hat{f}_\bk(\tau)\hat{f}^\dag_\bk(\tau')}}.
\end{equation}
The Fourier transformation of the Green's function and its inverse are respectively represented as
\begin{equation}
  \left\{
  \begin{aligned}
     & \gz(\bk,i\omega_n)=\frac{1}{2}\int^\beta_{-\beta}e^{i\omega_n\tau}\gz(\bk,\tau)d\tau, \\
     & \gz(\bk,\tau)=k_B T_{\mathrm{em}}\sum_ne^{-i\omega_n\tau}\gz(\bk,i\omega_n),
  \end{aligned}
  \right.
  \label{eq:fourier}
\end{equation}
where $k_B$ and $T_{\mathrm{em}}$ are Boltzmann constant and temperature, respectively.
$\omega_n=(2n+1)\pi k_BT_{\mathrm{em}} \; (n\in\mathbb{Z})$ is the fermionic Matsubara frequency.
Since the Hamiltonian $H_0$ is the fermion bilinear form, the Green's function can be easily computed as
$\gz(\bk,i\omega_n)=(i\omega_n-E_\bk)^{-1}$.

Substituting Eq.\autoref{eq:fourier} and $\gz(\bk,i\omega_n)$ to Eq.\autoref{eq:suscepImTime},
we arrive at the Fourier component of $\Phi_0(\tau,\tau')$,
\begin{align}
  \Phi_0(i\omega_n) & = \frac{k_B T_{\mathrm{em}}}{N}\sumk\sum_m\qty|p^{12}_\bk|^2\gz(\bk,i\omega_m)\qty(\gz(-\bk,i\omega_n-i\omega_m)+\gz(-\bk,-i\omega_n-i\omega_m))\notag              \\
                    & =\frac{k_B T_{\mathrm{em}}}{N}\sumk\sum_m\qty|p^{12}_\bk|^2\frac{1}{i\omega_m-E_\bk}\qty(\frac{1}{i\omega_n-i\omega_m-E_\bk}-\frac{1}{i\omega_n+i\omega_m+E_\bk}) .
  \label{eq:suscepFourier}
\end{align}

\begin{figure}[htb]
  \centering
  \includegraphics[width=0.55\linewidth]{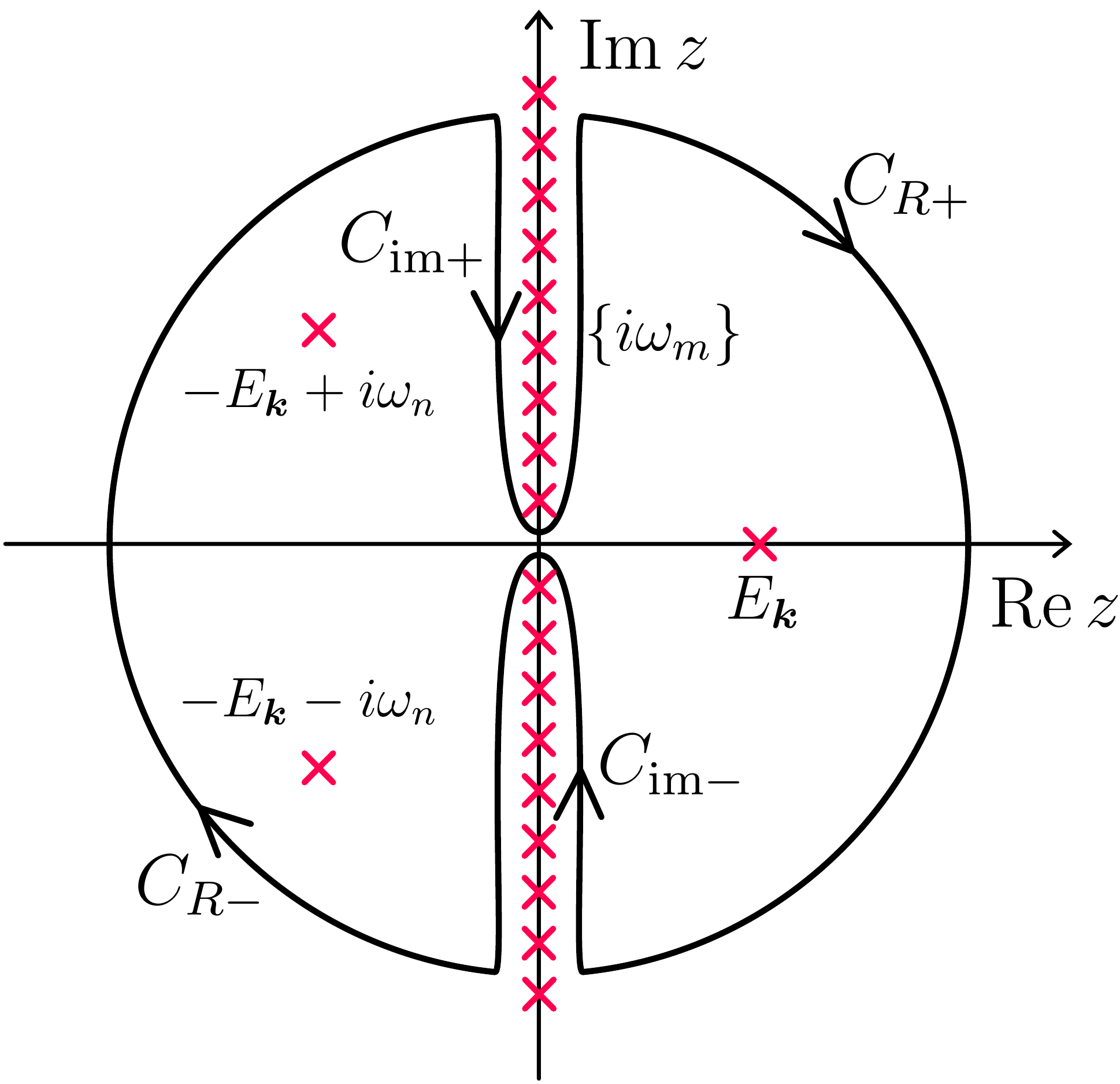}
  \caption{Integral paths and singularity points of $F(z)$ in the complex plane of $z$.}
  \label{fig:complexIntegral}
\end{figure}

To compute Eq.\autoref{eq:suscepFourier},
we apply the residue theorem and replace  the summation of $m$ with complex integrals.
We introduce a complex function $F(z)=F_1(z)F_2(z)$, where $F_1(z)$ and $F_2(z)$ are given by
\begin{equation}
  \left\{
  \begin{aligned}
     & F_1(z)=\frac{1}{e^{\beta z}+1},                                                        \\
     & F_2(z)=\frac{1}{z-E_\bk}\qty(\frac{1}{i\omega_n-z-E_\bk}-\frac{1}{i\omega_n+z+E_\bk}).
  \end{aligned}
  \right.
\end{equation}
Figure\autoref{fig:complexIntegral} shows the singularity points of $F(z)$ in the full complex plane of $z$.
$F_1(z)$ has singular points on the imaginary axis at
$i\omega_m=(2m+1)\pi k_BT_{\mathrm{em}} \; (n\in\mathbb{Z})$, and
$F_2(z)$ has three singular points at $(z_1,z_2,z_3)=(E_\bk,-E_\bk+i\omega_n,-E_\bk-i\omega_n)$.
From the residue theorem, we find the equality
\begin{equation}
  \qty(\oint_{C_{\mathrm{im-}}}+\oint_{C_{\mathrm{im+}}})\frac{dz}{2\pi i}F(z)=\sum_m\Res F(z=i\omega_m)=-k_B T_{\mathrm{em}}\sum_m\frac{1}{i\omega_m-E_\bk}\qty(\frac{1}{i\omega_n-i\omega_m-E_\bk}-\frac{1}{i\omega_n+i\omega_m+E_\bk}),
\end{equation}
where the paths, $C_{\mathrm{im-}}$ and $C_{\mathrm{im+}}$, respectively surround the imaginary axis
in the regime of ${\rm Im}z>0$ and ${\rm Im}z<0$, as shown in Fig.\autoref{fig:complexIntegral}.
$\Res F(a)$ denotes the residue of $F(z)$ at the point $z=a$.
Thus we can rewrite Eq.\autoref{eq:suscepFourier} as
\begin{equation}
  \Phi_0(i\omega_n) =-\frac{1}{N}\sumk\qty|p^{12}_\bk|^2\qty(\oint_{C_{\mathrm{im-}}}+\oint_{C_{\mathrm{im+}}})\frac{dz}{2\pi i}F(z).
  \label{eq:suscepInt1}
\end{equation}

Adding the paths $C_{R+}$ and $C_{R-}$ into $C_{\mathrm{im-}}$ and $C_{\mathrm{im+}}$
(see Fig.\autoref{fig:complexIntegral}),
we can rewrite Eq.\autoref{eq:suscepInt1} as
\begin{equation}
  \Phi_0(i\omega_n) =-\frac{1}{N}\sumk\qty|p^{12}_\bk|^2\qty(\int_{C_{\mathrm{im-}}} + \int_{C_{R-}}+\int_{C_{\mathrm{im+}}}+\int_{C_{R+}})\frac{dz}{2\pi i}F(z).
  \label{eq:suscepInt2}
\end{equation}
This operation is allowed because of $F(z)\sim z^{-2}$ at $z\rightarrow\infty$.
Applying the residue theorem to this closed path,
we find the equality
\begin{align}
  -\qty(\int_{C_{\mathrm{im-}}} + \int_{C_{R-}}+\int_{C_{\mathrm{im+}}}+\int_{C_{R+}})\frac{dz}{2\pi i}F(z) & =\sum_i^3\Res F(z=z_i)\notag                                                                                                \\
                                                                                                            & =\qty(\frac{1}{e^{\beta E_\bk}+1}+\frac{1}{e^{-\beta E_\bk}-1})\qty(\frac{1}{i\omega_n-2E_\bk}-\frac{1}{i\omega_n+2E_\bk}).
  \label{eq:complexInt}
\end{align}
Therefore, substituting Eq.\autoref{eq:complexInt} into Eq.\autoref{eq:suscepInt2}, we arrive at
\begin{equation}
  \Phi_0(i\omega_n) =\frac{1}{N}\sumk\qty|p^{12}_\bk|^2\qty(\frac{1}{e^{\beta E_\bk}+1}+\frac{1}{e^{-\beta E_\bk}-1})\qty(\frac{1}{i\omega_n-2E_\bk}-\frac{1}{i\omega_n+2E_\bk}).
\end{equation}

Finally, we perform the analytical continuation $i\omega_n\rightarrow \omega+i\delta$, which changes
$\Phi_0(i\omega_n)$ into $\chi(\omega)=\Phi_0(\omega+i\delta)$. The susceptibility is given by
\begin{equation}
  \chi(\omega)=\frac{1}{N}\sumk\qty|p^{12}_\bk|^2\qty(\frac{1}{e^{\beta E_\bk}+1}+\frac{1}{e^{-\beta E_\bk}-1})\qty(\frac{1}{\omega-2E_\bk+i\delta}-\frac{1}{\omega+2E_\bk+i\delta}).
\end{equation}
Taking the zero-temperature limit $\beta\rightarrow\infty\;(T_{\mathrm{em}}\rightarrow 0)$, we obtain
\begin{equation}
  \chi_P(\omega)=\lim_{\beta\rightarrow\infty}\chi(\omega)=\frac{1}{N}\sumk\qty|p^{12}_\bk|^2\qty(\frac{1}{\omega+2E_\bk+i\delta}-\frac{1}{\omega-2E_\bk+i\delta}).
\end{equation}

In our set up of high harmonic generation, we apply a THz laser pulse (not continuous wave) $E_\ac(t)=E_\ac\cos(\Omega t)f(t)$ with $f(t)=\exp[-2(\ln2)(t^2/t^2_{\mathrm{FWHM}})]$ to the Kitaev magnet.
Its Fourier component $E_\ac(\omega)$ is given by
\begin{equation}
  E_\ac(\omega)=\frac{E_\ac}{2}\sqrt{\frac{\pi}{A}}\qty(e^{-\frac{(\omega+\Omega)^2}{4A}}+e^{-\frac{(\omega-\Omega)^2}{4A}}),
\end{equation}
where $A=(2\log 2)/t^2_{\mathrm{FWHM}}$. In the Green's function, $1/\delta$ is viewed as the life time of the fermions. Therefore, to compare the result of our Lindblad equation with the linear response theory,
we set $\delta$ to be equal to the relaxation rate $\gamma$ of the Lindblad equation\autoref{eq:lindblad}.
In Figs.~\hyperref[fig:omegaExt-dep2d-123HG]{3} and~\hyperref[fig:kappa-dep]{4} of the main text, we have depicted both the results of the Lindblad equation and
the linear response theory.

\section{S6. SELECTION RULE for HHG SPECTRA}
\setcounter{section}{6}
\setcounter{equation}{0}
\label{sec:dynamicalSym}

In this section, we prove the characteristic selection rule~\cite{Alon1998,Morimoto2017,Neufeld2019} of high harmonic generation (HHG) in our Kitaev model.
This selection rule becomes exact when incident field is a continuous wave, i.e., $t_{\mathrm{FWHM}}\rightarrow\infty$.
Thus, in the following, we assume the continuous wave is applied to the Kitaev model, i.e., envelop function is $f(t)=1$.

The Hamiltonian for the isotropic Kitaev model
without both the dimerization $H_{\mathrm{ms}}$ and the three-spin interaction $H_\kappa$ has following symmetry:
\begin{equation}
  UH_K U^{\dag}=H_K,
\end{equation}
where the unitary operator $U$ is defined by
\begin{equation}
  U=U_{\mathrm{inv}}\Pi_z,
\end{equation}
with
\begin{align}
   & U_{\mathrm{inv}}\sigma_\br^{\alpha}U_{\mathrm{inv}}^{\dag}  =\sigma_{\br'}^{\alpha}, \\
   & \Pi_z\sigma_\br^{\alpha}\Pi_z^{\dag}                        =
  \left\{
  \begin{aligned}
     & \sigma_\br^y\quad(\alpha=x),  \\
     & -\sigma_\br^x\quad(\alpha=y), \\
     & \sigma_\br^z\quad(\alpha=z),
  \end{aligned}
  \right.
\end{align}
where $\br'=(-x,y)^\top$.
In the representation of Jordan--Wigner (JW) fermion, this symmetry is expressed as
\begin{equation}
  U d_\br U^\dag=d_{\br'}.
  \label{eq:JWconversion}
\end{equation}
Through Fourier transformation, Eq.\autoref{eq:JWconversion} is rewritten as
\begin{equation}
  U\tilde{d}_\bk U^\dag=\tilde{d}_{\bk'},
\end{equation}
where $\bk'=(-k_x,k_y)^\top$.
Thus, the Hamiltonian,
\begin{equation}
  H_K =\sumk \hat{h}(\bk) =\sumk\qty[\epsilon^{\mathrm{iso}}_\bk \tild_\bk^\dag \tild_\bk-\epsilon^{\mathrm{iso}}_\bk \tild_{-\bk} \tild_{-\bk}^\dag-i\Delta^{\mathrm{iso}}_\bk\qty(\tild_\bk^\dag \tild_{-\bk}^\dag-\tild_{-\bk}\tild_\bk)],
\end{equation}
satisfies the following transformation:
\begin{equation}
  U\hat{h}(\bk)U^\dag=\hat{h}(\bk').
  \label{eq:hamSym}
\end{equation}
Following the similar argument, we find that the polarization $\hat{P}=\sum'_\bk \hat{p}(\bk)$ obeys
\begin{equation}
  U\hat{p}(\bk)U^\dag=-\hat{p}(\bk').
  \label{eq:polSym}
\end{equation}
Two equations\autoref{eq:hamSym} and\autoref{eq:polSym} tell us that the driven Kitaev Hamiltonian
$H_K-E_\ac(t)\hat{P}=\sum'_\bk(\hat{h}(\bk)-E_\ac(t)\hat{p}(\bk))$ follows
\begin{equation}
  U\hat{h}(\bk,t)U^\dag=\hat{h}(\bk',t+T/2),
  \label{eq:dynamical-sym}
\end{equation}
where $\hat{h}(\bk,t)=\hat{h}(\bk)-E_\ac(t)\hat{p}(\bk)$.

Next, we consider a dynamical symmetry of the density matrix.
We symbolically represent the quantum master equation\autoref{eq:quantum-master-eq}
by using the Liouvillian super-operator $\mathcal{L}_\bk(t)$:
\begin{equation}
  \dv{t}\hat{\rho}(\bk,t)=\mathcal{L}_\bk(t)\hat{\rho}(\bk,t).
  \label{eq:quantum-master-eq}
\end{equation}
Our Lindblad operator satisfies
\begin{equation}
  UL_\bk U^\dag=e^{i\theta_\bk}L_{\bk'}\quad(\theta_\bk\in \mathbb{R}).
  \label{eq:lindbladSym}
\end{equation}
From two equations\autoref{eq:lindbladSym} and\autoref{eq:dynamical-sym}, we obtain
\begin{align}
  \dv{t}\hat{\rho}(\bk',t+T/2) & =\mathcal{L}_{\bk'}\hat{\rho}(\bk',t+T/2) 
  =U\mathcal{L}_\bk U^\dag\hat{\rho}(\bk',t+T/2),
\end{align}
hence
\begin{equation}
  \dv{t}U^\dag\hat{\rho}(\bk',t+T/2)U=\mathcal{L}_\bk U^\dag\hat{\rho}(\bk',t+T/2)U.
  \label{eq:dens-unitary}
\end{equation}
Comparing Eqs.\autoref{eq:quantum-master-eq} and\autoref{eq:dens-unitary}, we find the equality
\begin{equation}
  U\hat{\rho}(\bk,t)U^\dag=\hat{\rho}(\bk',t+T/2).
\end{equation}

Using this dynamical symmetry of the density matrix,
we compute the Fourier component of the polarization as
\begin{align}
  p(\bk,t) = \Braket{\hat{p}(\bk)}_t & =\Tr\qty[\hat{\rho}(\bk,t)\hat{p}(\bk)]  \notag       \\
                                     & =-\Tr\qty[\hat{\rho}(\bk',t+T/2)\hat{p}(\bk')] \notag \\
                                     & =-p(\bk',t+T/2).
\end{align}
This leads to
\begin{align}
  P(t) & =\frac{1}{N}\sumk p(\bk,t)   \notag    \\
       & =-\frac{1}{N}\sumk p(\bk',t+T/2)\notag \\
       & =-\frac{1}{N}\sumk p(\bk,t+T/2) \notag \\
       & =-P(t+T/2),
  \label{eq:polarization-periodicity}
\end{align}
where we have used $p(\bk,t)=p(-\bk,t)$ in the third line.
This result directly leads to the absence of the even-order HHG spectra.
Namely, from Eq.\autoref{eq:polarization-periodicity}, the $n$-th harmonic polarization
$P(n\Omega)$ is calculated as
\begin{align}
  P(n\Omega) & =\int_0^T\frac{dt}{T}P(t)e^{in\Omega t}\notag                                  \\
             & =\int_{-T/2}^{T/2}\frac{dt}{T}P(t+T/2)e^{in\Omega(t+T/2)}\notag                \\
             & =-e^{in\Omega\frac{T}{2}}\int_{-T/2}^{T/2}\frac{dt}{T}P(t)e^{in\Omega t}\notag \\
             & =-e^{in\pi}P(n\Omega),
  \label{eq:polOdd}
\end{align}
which means
\begin{equation}
  P(2m\Omega)=0\quad(m\in\mathbb{Z}).
\end{equation}
That is, we find that the even-order HHGs are all absent in the system of $H_K-E_\ac(t)\hat{P}$.
Our numerical results are consistent with this symmetry argument at $E_\dc=0$. See Fig.~\hyperref[fig:Estat-dep]{2}(a) of the main text.

Applying a DC electric field $E_\dc$ to the system, the Hamiltonian is modified to
\begin{equation}
  H(t)=H_0-E_\dc\hat{P}-E_{\mathrm{ac}}(t)\hat{P}.
\end{equation}
In this case, the dynamical symmetry is broken and
\begin{equation}
  UH(t)U^\dag\neq H(t+T/2).
\end{equation}
Therefore, in a dimerized (i.e., anisotropic) Kitaev model,
even-order polarizations $P(2m\Omega)$ are generally not prohibited.
Our numerical calculation shows that a finite $P(2m\Omega)$ emerges if we introduce $E_\dc\neq 0$.

Next, we also consider the isotropic Kitaev model with a three-spin interaction term,
$H_K+H_\kappa$ with a finite $\kappa$.
In this case, let us introduce the time-reversal operator $\hat{V}$, which satisfies
\begin{equation}
  \hat{V}\sigma^\alpha_\br\hat{V}^\dag=-\sigma^\alpha_\br\quad(\alpha=x,y,z).
\end{equation}
Combining $\hat U$ and $\hat V$, we can find that the Hamiltonian $H_K+H_\kappa$ has following symmetry:
\begin{equation}
  \hat{V}\hat{U}(H_K+H_\kappa)(\hat{V}\hat{U})^\dag=H_K+H_\kappa.
\end{equation}
Replacing $\hat U$ with $\hat{V}\hat{U}$ in the above dynamical-symmetry argument about $H_K$,
we can show that even-order HHGs all disappear in the system of $H_K+H_\kappa-E_\ac(t)\hat{P}$.

\section{S7. MAGNETIC ANISOTROPY DEPENDENCE of EVEN-ORDER HARMONICS}
\setcounter{section}{7}
\setcounter{equation}{0}

\begin{figure}[htbp]
    \centering
    \includegraphics[width=\linewidth]{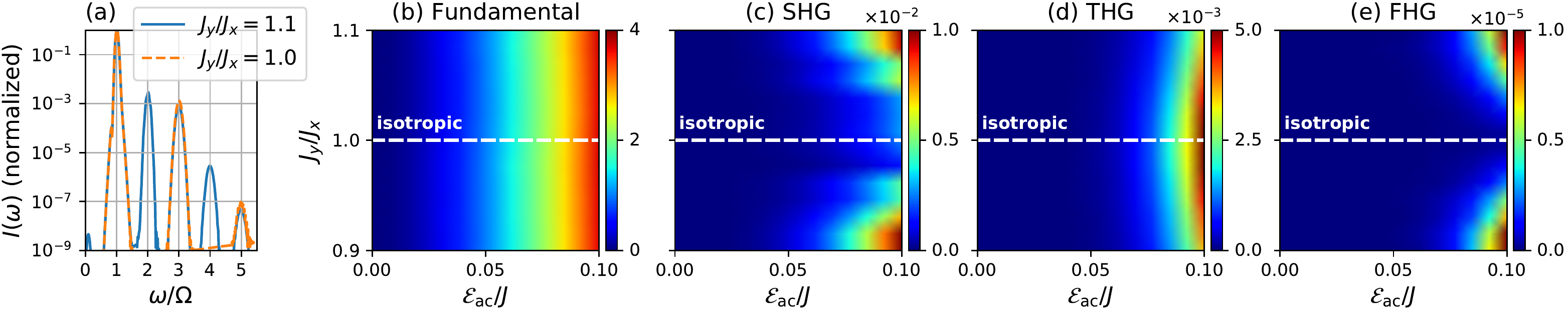}
    \caption{
    HHG spectra $I(\omega)$ in anisotropic ferromagnetic ($J_{x,y,z}>0$) Kitaev models under a THz pulse of
    $\Omega=2.0J_x$. (a) $I(\omega)$ as a function of $\omega$ at $J_y/J_x=1$ and $1.1$ under the irradiation
    of $\eac=0.1J_x$. $I(\omega)$ is normalized with its maximum value. (b)--(e) $(\eac,J_y/J_x)$ dependence of
    fundamental harmonic [$I(\Omega)$], SHG [$I(2\Omega)$], THG [$I(3\Omega)$], and FHG [$I(4\Omega)$]
    spectra. The unit of intensity is chosen as $I(\Omega)$ at $\eac=0.05J_x$ and $J_y/J_x=1$.
    The other parameters are set to be $J_z=J_x$, $\gamma=0.1J_x$, and $\edc=0$.}
    \label{fig:Anisotropy}
\end{figure}

As discussed in the main text and the above section,
we can apply the dynamical symmetry and selection rule in a wide class of Kitaev models with various perturbations.
In the main text, we have focused on the isotropic point ($J_{x,y,z}=J$) while Kitaev candidates sometimes possess a magnetic anisotropy, i.e., inequivalence of $J_x$, $J_y$, and $J_z$.
Here, we study the even-order harmonics in an anisotropic Kitaev model without dc electric field.
The argument in the above section tells us that the dynamical symmetry survives for $J_z\neq J_x=J_y$,
while it breaks down for $J_x\neq J_y$. Therefore, the anisotropic Kitaev models with $J_x\neq J_y$ are
expected to possess a finite intensity of even-order harmonics.
Using the master equation, we numerically calculate the anisotropy dependence of the even-order harmonics
in the anisotropic Kitaev models and the result is given in Fig.\autoref{fig:Anisotropy}.
This figure clearly indicates that a finite anisotropy ($J_x\neq J_y$) indeed induces peaks of even-order harmonics
(SHG and FHG).

\section{S8. \texorpdfstring{$\kappa$}{KAPPA} DEPENDENCE of DENSITY of STATES}
\setcounter{section}{8}
\setcounter{equation}{0}

\begin{figure}[htb]
  \centering
  \includegraphics[width=\linewidth]{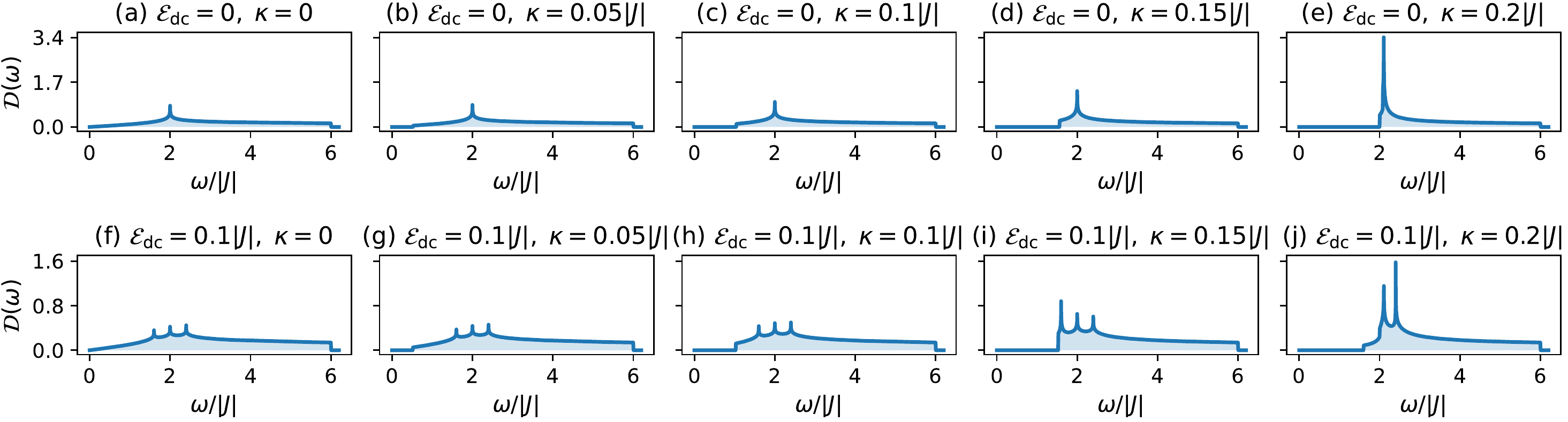}
  \caption{$\kappa$ dependences of DoSs in the antiferromagnetic ($J<0$) Kitaev models
    at $\kappa/\abs{J}=$ (0, 0.05, 0.1, 0.15, 0.2) with (a--e) $\edc=0$ and (f--j) $\edc=0.1\abs{J}$.}
  \label{fig:DoS-kappa}
\end{figure}

\begin{figure}[htb]
  \centering
  \includegraphics[width=\linewidth]{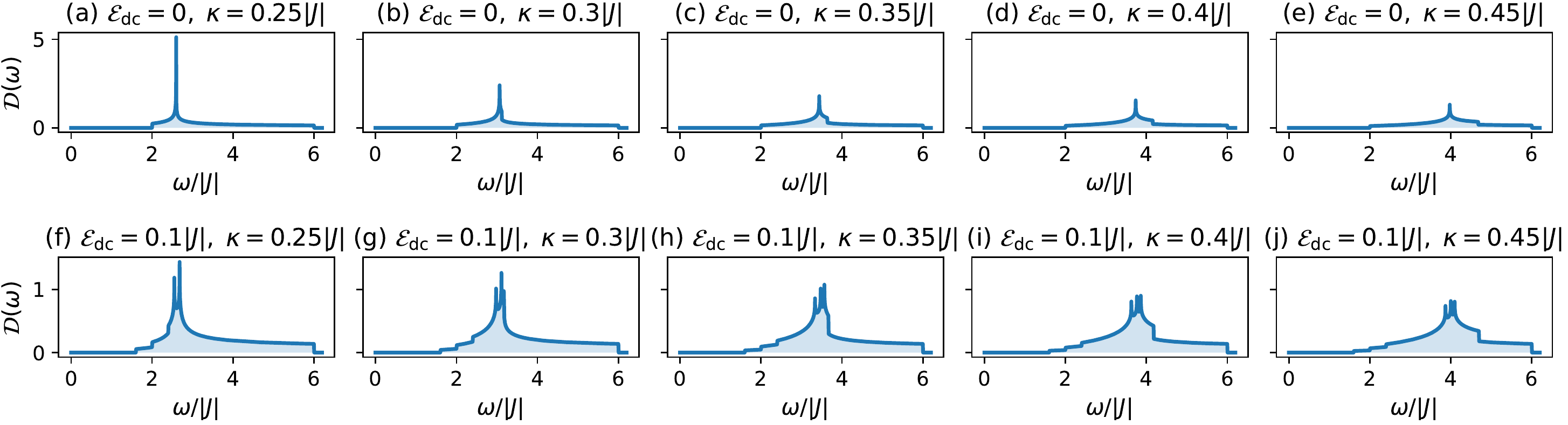}
  \caption{$\kappa$ dependences of DoSs in the antiferromagnetic ($J<0$) Kitaev models
    at $\kappa/\abs{J}=$ (0.25, 0.3, 0.35, 0.4, 0.45) with (a--e) $\edc=0$ and (f--j) $\edc=0.1\abs{J}$.}
  \label{fig:DoS-kappa-high}
\end{figure}

In the main text, we have discussed the $\kappa$ dependence of the density of state (DoS)
$\mathcal{D}(\omega)$ of fermions $f_\bk$ in the antiferromagnetic Kitaev model with $J<0$.
Physically, $\kappa$ is proportional to the cubic of external magnetic field $\bm{B}$.
Therefore, the $\kappa$-dependent DoS tells us its $\abs{\bm{B}}$ dependence.
In this section, we shortly discuss the shape of $\mathcal{D}(\omega)$ as a function of $\kappa$ in more detail.
Figures\autoref{fig:DoS-kappa} and\autoref{fig:DoS-kappa-high} depict
DoSs $\mathcal{D}(\omega)$ for different value of $\kappa$.
Panels (a)--(e) are the result for the isotropic Kitaev models with $E_\dc=0$,
while panels (f)--(j) are those of the dimerized model with $E_\dc\eta_{\mathrm{ms}}=0.1 \abs{J}$.

In Fig.\autoref{fig:DoS-kappa}, we focus on the weak $\kappa$ region of $\kappa\alt 0.2 \abs{J}$,
which correspond to the regime of weak dc magnetic fields.
The fermion mass gap $\Delta_\kappa$ monotonically increases with the growth of $\kappa\sim \abs{\bm{B}}^3$.
Panels (a)--(e) show that when the mass gap $\Delta_\kappa$ increases for $E_\dc=0$,
the highest value (peak height) of $\mathcal{D}(\Omega_0)$ i.e., $\mathcal{D}(\Omega_0\sim 2\abs{J})$,
also grows up.
For $E_\dc\neq 0$, the peak of $\mathcal{D}(\Omega_0)$ is split into three peaks, as shown in panels (f)--(j).
However, a similar $\kappa$ dependence remains, i.e., the height of three peaks increases
with the growth of $\Delta_\kappa$. We note that the peak frequency $\Omega_0\sim 2|J|$
is almost unchanged in this weak $\kappa$ regime.

On the other hand, Fig.\autoref{fig:DoS-kappa-high} tells us that
different behavior of the gap $\Delta_\kappa$ and peak frequency appears
in the strong $\kappa$ regime of $\kappa\agt 0.2\abs{J}$.
Figure\autoref{fig:DoS-kappa-high} shows
that the mass gap $\Delta_\kappa$ is almost fixed at $\Delta_\kappa\sim2\abs{J}$,
while the peak position $\Omega_0$ of $\mathcal{D}(\omega)$ almost linearly increases with
$\kappa\sim\abs{\bm{B}}^3$ for both $E_\dc=0$ and $E_\dc\neq0$.

\section{S9. HHG in ANTIFERROMAGNETIC KITAEV MODEL with DC ELECTRIC FIELD}
\setcounter{section}{9}
\setcounter{equation}{0}

In the main text, we have shown some characteristic features of the dc-electric-field dependence of HHG
in the ferromagnetic Kitaev model. In this section, we discuss the dc-electric-field $E_\dc$ and
laser-frequency $\Omega$ dependences of HHG in antiferromagnetic Kitaev model $(J<0)$.

\begin{figure}[htb]
  \centering
  \includegraphics[width=\linewidth]{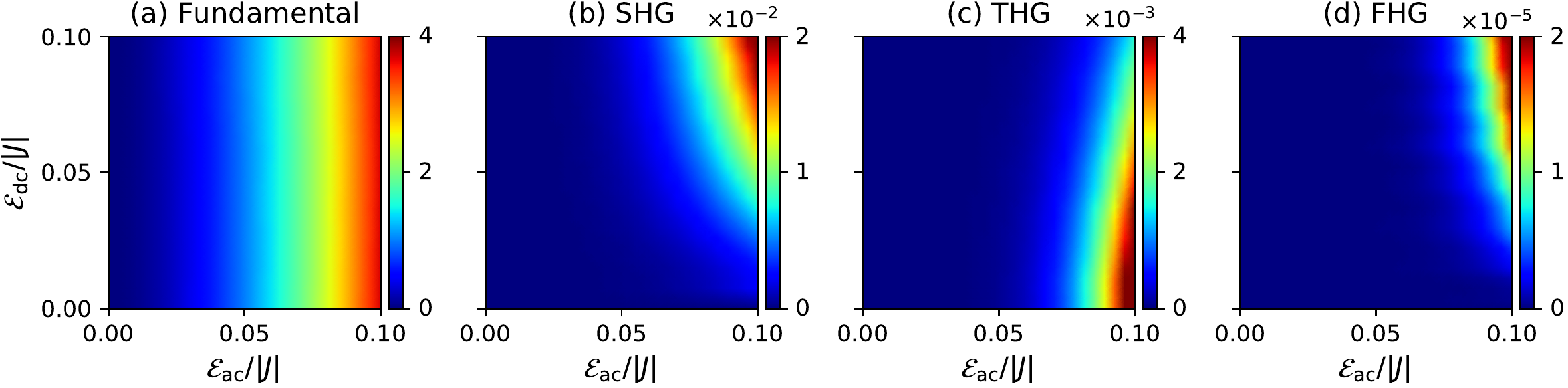}
  \caption{Intensities of (a) fundamental harmonic [$I(\Omega)$], (b) SHG [$I(2\Omega)$], (c) THG [$I(3\Omega)$], and (d) FHG [$I(4\Omega)$] in the antiferromagnetic ($J<0$) Kitaev models driven by THz laser of $\Omega=2.0\abs{J}$ in the space $(\eac,\edc)$. The unit of intensity is chosen as $I(\Omega)$ at $\eac=0.05\abs{J}$ and $\edc=0.05\abs{J}$. The relaxation rate is set to be $\gamma=0.1\abs{J}$.}
  \label{fig:antiFerroEs}
\end{figure}

Figure~\ref{fig:antiFerroEs} represents the $(E_\ac,E_\dc)$ dependences of fundamental harmonic,
SHG, THG, and FHG in the antiferromagnetic Kitaev models. From the panels (b) and (d), we see that even-order harmonics (SHG and FHG) appear only when a finite dc field $E_\dc$ is introduced to the system.
This is owing to the fact that there is the same dynamical symmetry as the ferromagnetic Kitaev model [Sec.~\hyperref[sec:dynamicalSym]{S6}] for the case of zero dc electric field $E_\dc=0$.
We have verified that the $(E_\ac,E_\dc)$ dependences of HHG in the antiferromagnetic case are
very similar to those in the ferromagnetic one.

\begin{figure}[htb]
  \centering
  \includegraphics[width=0.8\linewidth]{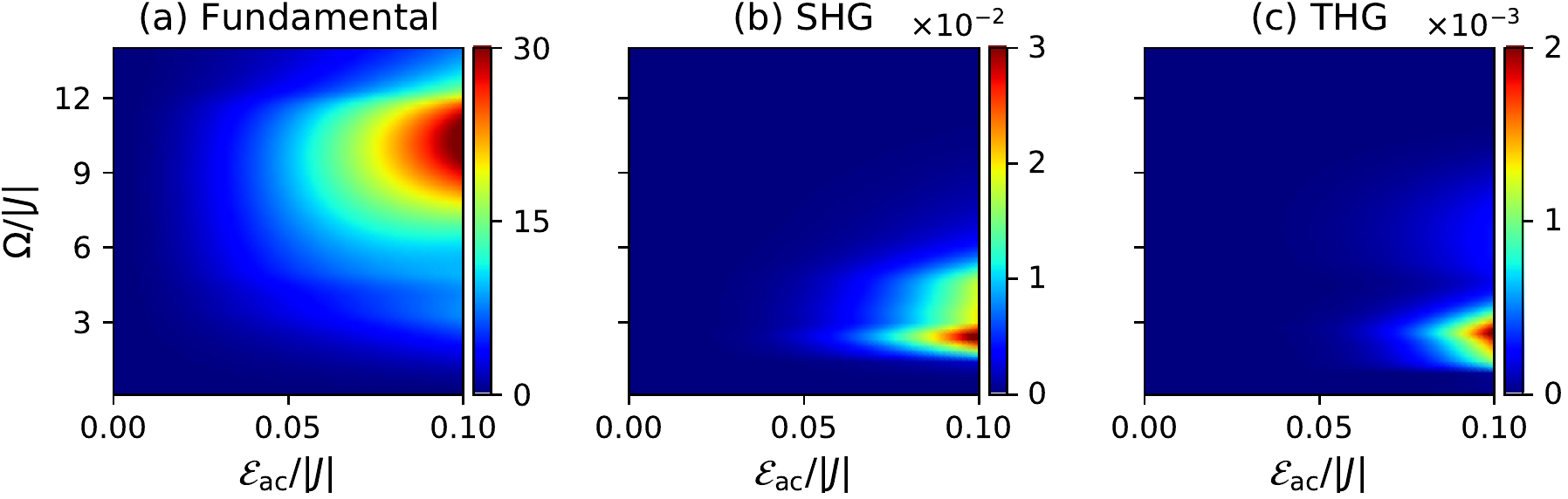}
  \caption{Intensities of (a) fundamental harmonic [$I(\Omega)$], (b) SHG [$I(2\Omega)$], and (c) THG [$I(3\Omega)$] in the driven antiferromagnetic ($J<0$) Kitaev models with laser of $\Omega=2\abs{J}$ and $\edc=0.1\abs{J}$
    in the space $(\eac,\Omega)$. The unit of intensity is chosen as $I(\Omega)$ at $\eac=0.05\abs{J}$ and $\Omega=2.0\abs{J}$. The relaxation rate is $\gamma=0.1\abs{J}$.}
  \label{fig:antiFerroOmega}
\end{figure}

Figure~\ref{fig:antiFerroOmega} shows a fundamental harmonic, SHG, and THG of
the same antiferromagnetic Kitaev models in the space $(E_\ac,\Omega)$.
We find broad peaks in all of $I(\Omega)$, $I(2\Omega)$, and $I(3\Omega)$.
The peaks of $I(2\Omega)$ and $I(3\Omega)$ appear around $\Omega=\Omega_{\mathrm{peak}}/2$ and
$\Omega_{\mathrm{peak}}/3$, respectively. This is the natural result
from the view of the perturbation theory with respect to the laser intensity.
These results are also similar to those of the ferromagnetic model.

\section{S10. HHG in KITAEV HONEYCOMB MODEL with DC MAGNETIC FIELD}
\setcounter{section}{10}
\setcounter{equation}{0}

In this section, we discuss effect of a dc magnetic field on the HHG of the Kitaev model.
In our present study, the effect of dc magnetic field is taken in the $\kappa$ term $H_\kappa$
of the three-spin interaction. As we already mentioned, the microscopic origin of
the coupling constant $\kappa$ is a usual Zeeman coupling with dc magnetic field $\bm{B}$
and the relation $\kappa\sim \abs{\bm{B}}^3$ is hold. As we mentioned in the main text,
when we apply a dc magnetic field $\bm{B}$ to a static Kitaev model,
the spin liquid state suddenly changes into a ferromagnetic phase for ferromagnetic case ($J>0$).
On the other hand, the spin liquid phase is relatively stable against $\bm{B}$
in the antiferromagnetic model with $J<0$.
Therefore, we note that the antiferromagnetic Kitaev model is more realistic
if we take into account the effect of $\bm{B}$ by introducing the $\kappa$ term.

\begin{figure}[htb]
  \centering
  \includegraphics[width=\linewidth]{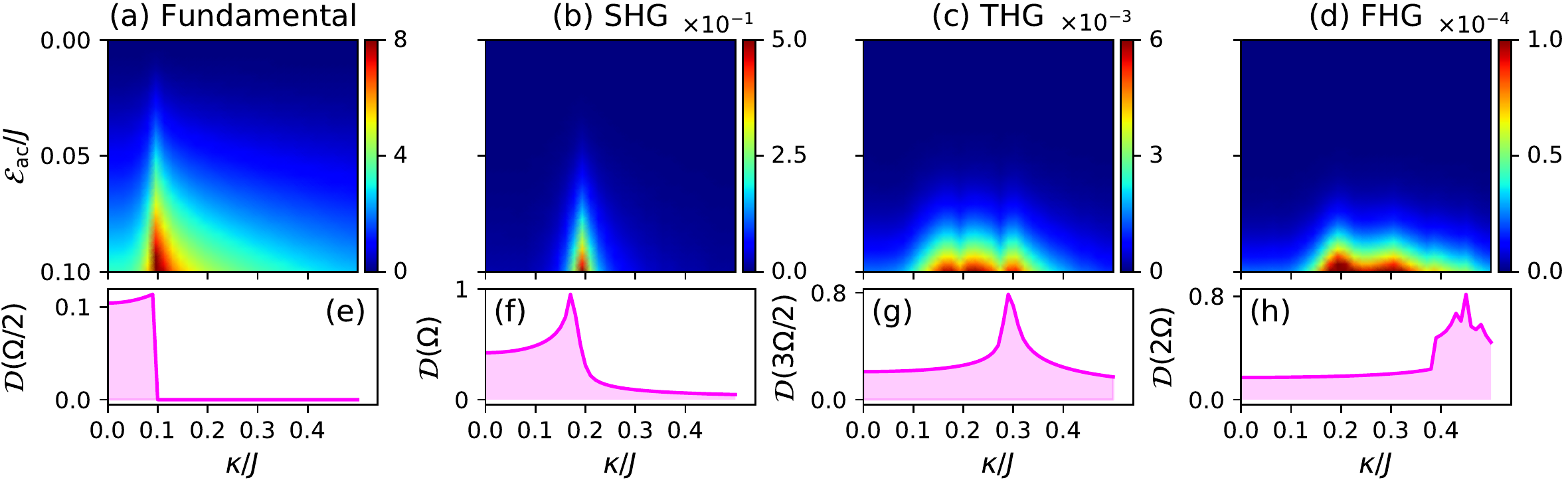}
  \caption{Intensities of (a) fundamental harmonic [$I(\Omega)$], (b) SHG [$I(2\Omega)$], (c) THG [$I(3\Omega)$], and (d) FHG [$I(4\Omega)$] in the slightly-anisotropic ($\edc=0.1J$) ferromagnetic ($J>0$) Kitaev models
    driven by laser pulse of $\Omega=2J$ in the space $(\eac,\kappa)$.
    (e--h) DoS $\mathcal{D}(\omega)$ as a function of $\kappa$: (e)--(h) respectively correspond to panels (a)--(d).
    The unit of intensity is chosen as $I(\Omega)$ at $\eac=0.05J$ and $\kappa=0.05J$.
    The relaxation rate is set to be $\gamma=0.1J$.}
  \label{fig:ferroKappa}
\end{figure}

\begin{figure}[htb]
  \centering
  \includegraphics[width=\linewidth]{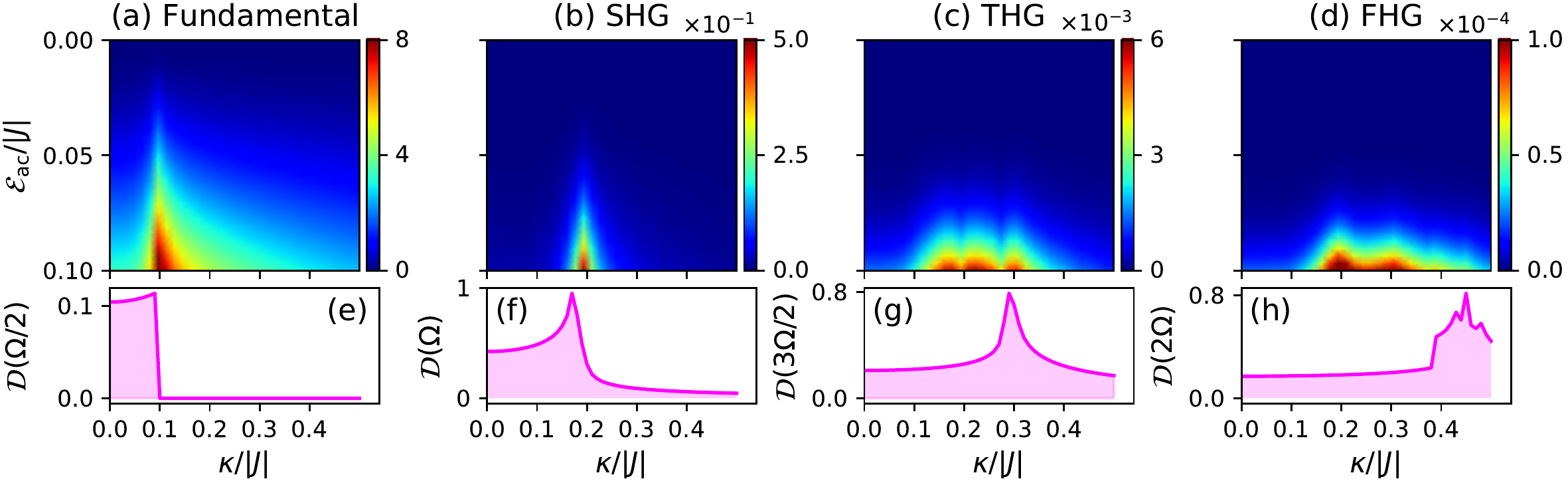}
  \caption{Intensities of (a) fundamental harmonic [$I(\Omega)$], (b) SHG [$I(2\Omega)$], (c) THG [$I(3\Omega)$],
    and (d) FHG [$I(4\Omega)$] in the slightly-anisotropic ($\edc=0.1\abs{J}$) antiferromagnetic ($J<0$) Kitaev models
    driven by laser pulse of $\Omega=2\abs{J}$ in the space $(\eac,\kappa)$.
    (e--h) DoS $\mathcal{D}(\omega)$ as a function of $\kappa$: (e)--(h) respectively correspond to panels (a)--(d).
    The unit of intensity is chosen as $I(\Omega)$ at $\eac=0.05\abs{J}$ and $\kappa=0.05\abs{J}$.
    The relaxation rate is set to be $\gamma=0.1\abs{J}$.}
  \label{fig:antiferroKappa}
\end{figure}

First, we discuss the $(E_\ac,\kappa)$ dependences of fundamental harmonic, SHG, THG, and FHG
in ferromagnetic $(J>0)$ [Figs.\autoref{fig:ferroKappa}(a--d)] and antiferromagnetic $(J<0)$ [Figs.\autoref{fig:antiferroKappa}(a--d)] Kitaev models.
Panels (e--h) respectively show the $\kappa$ dependences of the fermion DoSs $\mathcal{D}(\omega)$
corresponding to panels (a--d).
The intensity peaks of the fundamental harmonic (SHG) appear when the laser frequency $\Omega$ approaches
the points where DoS $\mathcal{D}(\Omega/2)$ ($\mathcal{D}(\Omega)$) take the highest value.
This is natural from the perturbative picture of the laser effect.
However, the nontrivial peak distributions occur in the THG and FHG, and they cannot be explained
from the simple perturbation theory.
These peaks may be attributed to not only inter-band fermion-pair transition process
but also the intra-band dynamics, which will be explained in Sec.~\hyperref[sec:intra]{S11}.
We also find that there is no significant difference between the ferromagnetic and antiferromagnetic cases.

\begin{figure}[htb]
  \centering
  \includegraphics[width=0.85\linewidth]{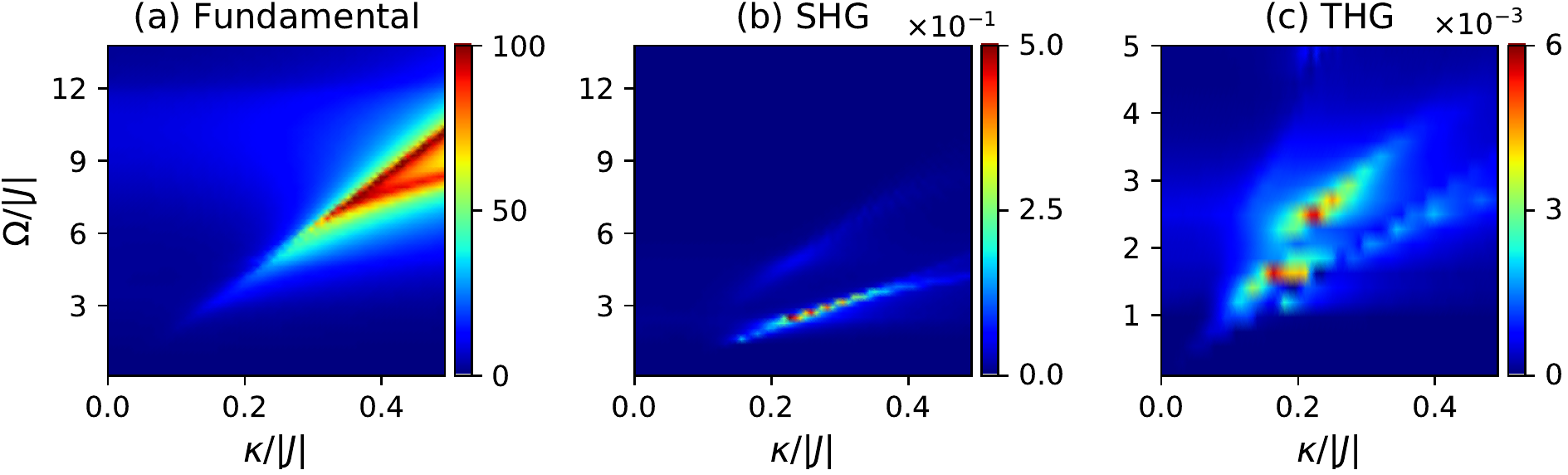}
  \caption{Intensities of (a) fundamental harmonic [$I(\Omega)$], (b) SHG [$I(2\Omega)$], and
    (c) THG [$I(3\Omega)$] in the slightly-anisotropic ($\edc=0.1\abs{J}$) antiferromagnetic ($J<0$) Kitaev models
    driven by laser pulse of intensity $\eac=0.1\abs{J}$ in the space $(\kappa,\Omega)$.
    The unit of intensity is chosen as $I(\Omega)$ at $\Omega=2.0\abs{J}$ and $\kappa=0.05\abs{J}$.
    The relaxation rate is set to be $\gamma=0.1\abs{J}$.}
  \label{fig:kappa-omega}
\end{figure}

\begin{figure}[htbp]
    \centering
    \includegraphics[width=\linewidth]{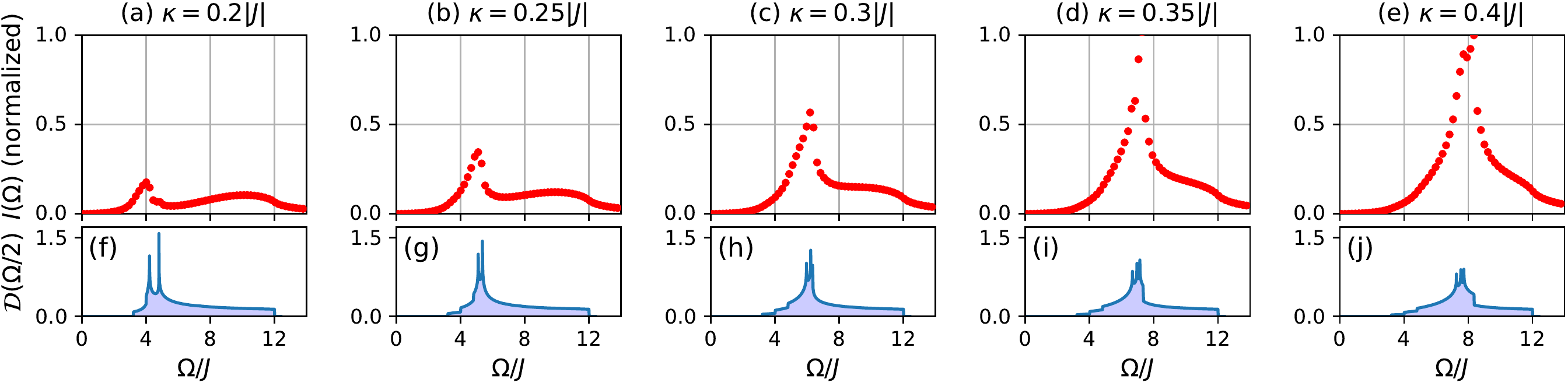}
    \caption{
    (a)--(e) $I(\Omega)$ of the antiferromagnetic ($J<0$) Kitaev models with $\edc=0.1\abs{J}$ and $\eac=0.1\abs{J}$ at (a) $\kappa/\abs{J}=0.2$, (b) $0.25$, (c) $0.3$, (d) $0.35$, and (e) $0.4$.
	$I(\Omega)$ in panels (a)--(e) are normalized with the maximum value in the panel (e).
	Panels (f)--(j) show DoSs of fermions corresponding to the cases (a)--(e), respectively.
    }
    \label{fig:depKappa}
\end{figure}

Next, in Fig.\autoref{fig:kappa-omega},
we depict the $(\kappa,\Omega)$ dependences of fundamental harmonic, SHG, and THG
in the antiferromagnetic Kitaev models $(J<0)$ under a strong laser pulse with $\eac=0.1\abs{J}$.
Figures\autoref{fig:kappa-omega}(a) and\autoref{fig:kappa-omega}(b) are also depicted in Fig.~\hyperref[fig:kappa-dep]{4} of the main text.
As we discussed in the main text,
the frequency $\Omega$ at the broad peak position monotonically increases
in an almost $\kappa$-linear fashion for fundamental harmonic $I(\Omega)$ in the case of $\kappa\agt 0.2\abs{J}$.
This is because the peak position $\Omega_0$ of $\mathcal{D}(\Omega)$
almost linearly increases with $\kappa$ for $\kappa\agt 0.2\abs{J}$.
We give Fig.\autoref{fig:depKappa} to more clearly show this relation among $\kappa$,
the peak position $\Omega_0$ of the DoS, and that of the fundamental harmonic $I(\Omega)$
(see also Fig.\autoref{fig:DoS-kappa-high}). One can verify that the peak position of $I(\Omega)$ indeed
coincides with that of the DoS, $\Omega_0$ for  $\kappa\agt 0.2\abs{J}$.
Since $\kappa\sim \abs{\bm{B}}^3$, the $\bm{B}$-cube dependent peak frequency is specific for the Kitaev model.
In usual magnets, the magnetic-resonance frequency $\Omega$ is proportional to $\abs{\bm{B}}$.
The peak frequency of $I(2\Omega)$ is almost half of $\Omega_{\mathrm{peak}}$ of $I(\Omega)$
as shown in Fig.\autoref{fig:kappa-omega}(b), and this is a natural result from the perturbative viewpoint.
However, the nontrivial peak occurs in the THG and this peak may be explained by
the non-perturbative intra-band dynamics. This intra-band dynamics will be discussed in Sec.~\hyperref[sec:intra]{S11}.

\begin{figure}[H]
  \centering
  \includegraphics[width=0.35\linewidth]{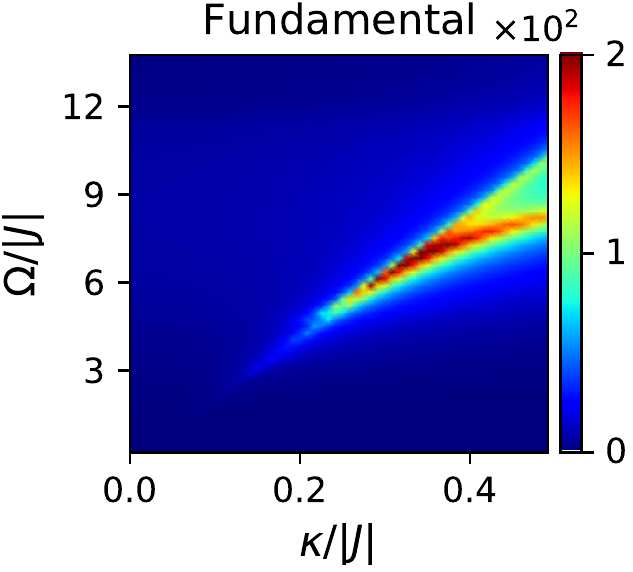}
  \caption{Intensity of fundamental harmonic [$I(\Omega)$] in the slightly-anisotropic ($\edc=0.1\abs{J}$) antiferromagnetic ($J<0$) Kitaev models
  driven by laser pulse of intensity $\eac=10^{-3}\abs{J}$ in the space $(\kappa,\Omega)$.
  The unit of intensity is chosen as $I(\Omega)$ at $\Omega=2.0\abs{J}$ and $\kappa=0.05\abs{J}$.
  The relaxation rate is set to be $\gamma=0.1\abs{J}$.}
  \label{fig:AFomegaKappa_weak}
\end{figure}

In the case of a weak pulse $(\eac=10^{-3}\abs{J})$,
we also observe a sharp peak of $I(\Omega)$ at $\Omega_{\mathrm{peak}}$ in Fig.\autoref{fig:AFomegaKappa_weak}.
This result indicates that,
even in the linear response regime, the fundamental harmonic shows characteristics of the Kitaev QSL
for $\kappa\agt0.2\abs{J}$.
$I(\Omega)$ driven by intense pulses with a finite magnetic field tells us the peak position of the
DoS $\Omega_0$, which is half as large as $\Omega_{\mathrm{peak}}$.

\section{S11. INTRA- and INTER-BAND DYNAMICS in DRIVEN KITAEV MODEL}
\setcounter{section}{11}
\setcounter{equation}{0}
\label{sec:intra}

In this section, we consider the laser-induced dynamics in the driven Kitaev model
from the microscopic viewpoint of the fermion band structure.
If the applied THz laser is sufficiently weak, the vacuum state is weakly perturbed by photons
and only the dynamics including a few fermions is important.
In this case, the resonant-like process creating fermion pairs with $\bk$ and $-\bk$ is dominant.
Namely, when the fermion-pair excitation energy coincides with the photon energy $\hbar\Omega$,
the fermion pairs are strongly created. This may be called the inter-band dynamics.
However, if the laser pulse becomes strong,
the situation is expected to be different.
In this case, in addition to the resonant pair creations and their relaxation,
the intra-band dynamics processes may also contribute to HHG
due to the nonlinear optical (multiple photon) effect.
We discuss this expectation from the numerical calculation below.

\begin{figure}[htb]
  \centering
  \includegraphics[width=\linewidth]{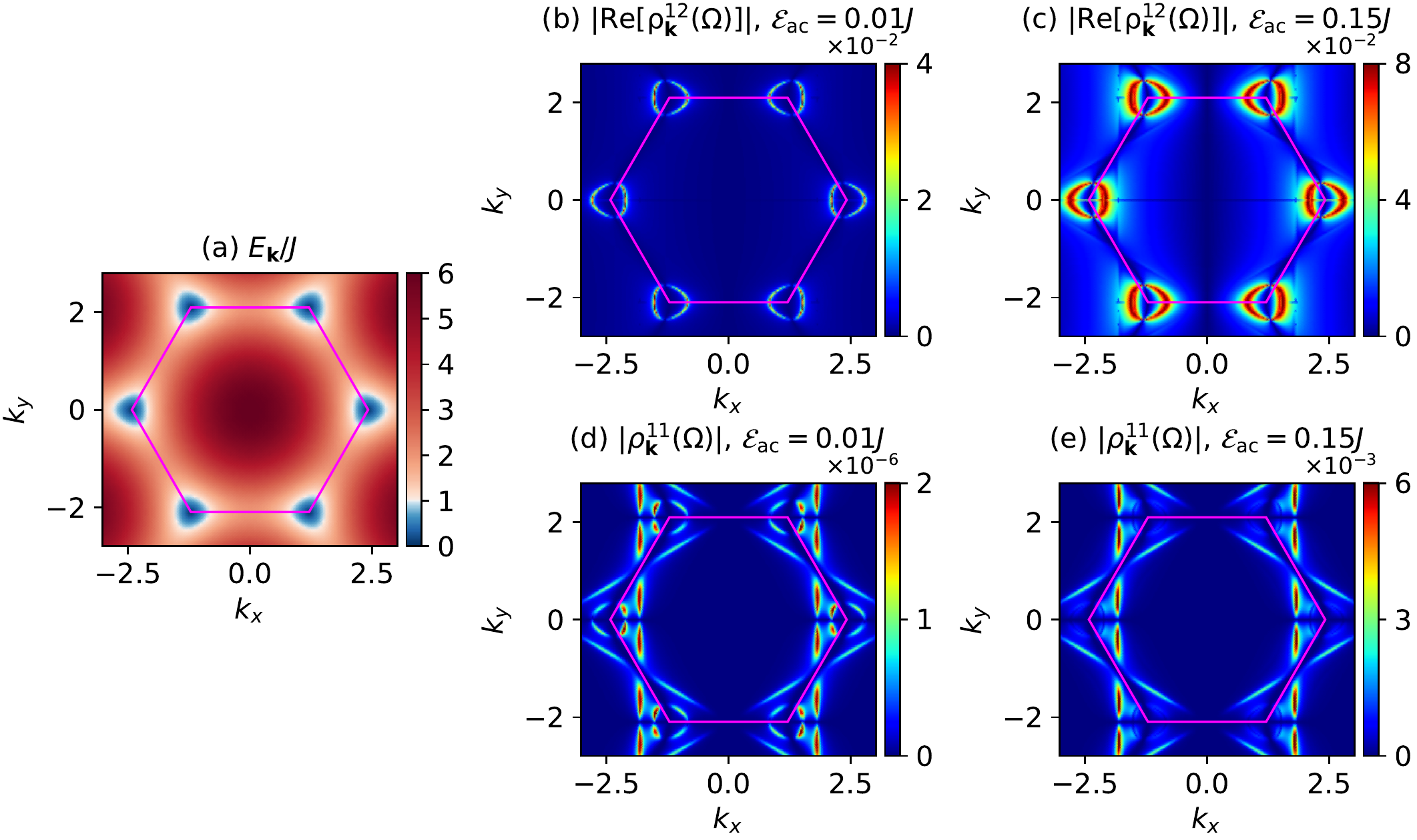}
  \caption{(a) Fermion band structure of the isotropic antiferromagnetic ($J<0$) Kitaev model $H_K$ with $\edc=\kappa=0$.
  The white areas around $K$ and $K'$ points in Brillouin zone correspond to the excitation energy $J$.
  (b),(c) Fourier components of the off-diagonal elements $\abs{\Re[\rho_{\bk}^{12}(\Omega)]}$
  for the Kitaev model driven by THz laser with frequency $\Omega=2J$ at (b) $\eac=10^{-2}J$ and (c) $\eac=0.15J$.
  (d),(e) Diagonal parts $\abs{\rho_{\bk}^{11}(\Omega)}$ for the same driven Kitaev model at
  (d) $\eac=10^{-2}J$ and (e) $\eac=0.15J$.
  We set the parameters of the Kitaev model to be $\edc=\kappa=0$,
  and the relaxation rate is set to be $\gamma=0.1J$.
  }
  \label{fig:interIntra}
\end{figure}

We have defined the $2\times 2$ density matrix $\rho_{\bk}$ with the bases of the ground state $\ket{g_\bk}$
and the excited one $\ket{e_\bk}$.
Therefore, the inter-band dynamics is reflected in the off-diagonal elements
$\rho_{\bk}^{12}=(\rho_{\bk}^{21})^*$,
while the intra-band dynamics can be observed in the diagonal part $\rho_{\bk}^{11}=1-\rho_{\bk}^{22}$.
We calculate the laser-driven dynamics of the density matrix $\rho_{\bk}$ in the full Brillouin zone.
Figure\autoref{fig:interIntra} shows the Fourier component of
$\abs{\rho_{\bk}^{11}}$ and $\abs{\Re[\rho_{\bk}^{12}]}$ for the Kitaev model under the laser with $\Omega=2J$.
As the reference, we show the Fermion energy band structure in Fig.\autoref{fig:interIntra}(a),
in which the regime of $\Omega/2=E_\bk/J=1$ is depicted by white color.
Figures\autoref{fig:interIntra}(b) and\autoref{fig:interIntra}(c) represent
the Fourier components of the off-diagonal elements $\abs{\Re[\rho_{\bk}^{12}(\Omega)]}$ for
weak ($\eac=10^{-2}J$) and strong ($\eac=0.15J$) laser intensities, respectively.
Similarly, Figs.\autoref{fig:interIntra}(d) and\autoref{fig:interIntra}(e) respectively show the off-diagonal
components $\abs{\rho_{\bk}^{11}(\Omega)}$ for weak and strong lasers.
From Figs.\autoref{fig:interIntra}(a)--\ref{fig:interIntra}(c), we see that the inter-band dynamics
(i.e., $\abs{\Re[\rho_{\bk}^{12}(\Omega)]}$) are much active around the regime of $\Omega/2=E_\bk/J=1$
irrespectively of the strength of laser.
It indicates that (as expected) the resonant-like dynamics is always dominant in HHG of the Kitaev model.

For weak laser pulse, Figs.\autoref{fig:interIntra}(b) and\autoref{fig:interIntra}(d) show that the ratio
$C(\eac=10^{-2}J)=\max[\abs{\rho_{\bk}^{11}(\Omega)}]/\max[\abs{\Re[\rho_{\bk}^{12}(\Omega)]}]\sim 10^{-4}$,
which indicates that the resonant-like process of the off-diagonal dynamics is dominant in the HHG.
(Here, $\max[\cdots]$ means the maximum value in the Brillouin zone.)
On the other hand, Figs.\autoref{fig:interIntra}(c) and\autoref{fig:interIntra}(e) show that the ratio increases as
$C(\eac=0.15J)\sim 10^{-1}$ for the strong THz pulse.
This means that the intra-band transition process becomes more important than that for a weak THz pulse.
The increase of $\abs{\rho_{\bk}^{11}(\Omega)}$ is thereby viewed as a signature of nonlinear optical effects.

From Figs.\autoref{fig:interIntra}(b)--(e), we also find that the regime with high value of
$\abs{\Re[\rho_{\bk}^{12}(\Omega)]}$ does not overlap well that with intense $\abs{\rho_{\bk}^{11}(\Omega)}$,
namely, the active areas of $\abs{\Re[\rho_{\bk}^{12}(\Omega)]}$ and $\abs{\rho_{\bk}^{11}(\Omega)}$
are independently of each other.

\section{S12. NONLINEARITY of HHG and REQUIRED LASER INTENSITY}
\setcounter{section}{12}
\setcounter{equation}{0}

In this section, we discuss the crossover from linear to nonlinear responses,
and estimate the laser intensity required to observe HHG in the experiment.

\begin{figure}[htb]
  \centering
  \includegraphics[width=0.5\linewidth]{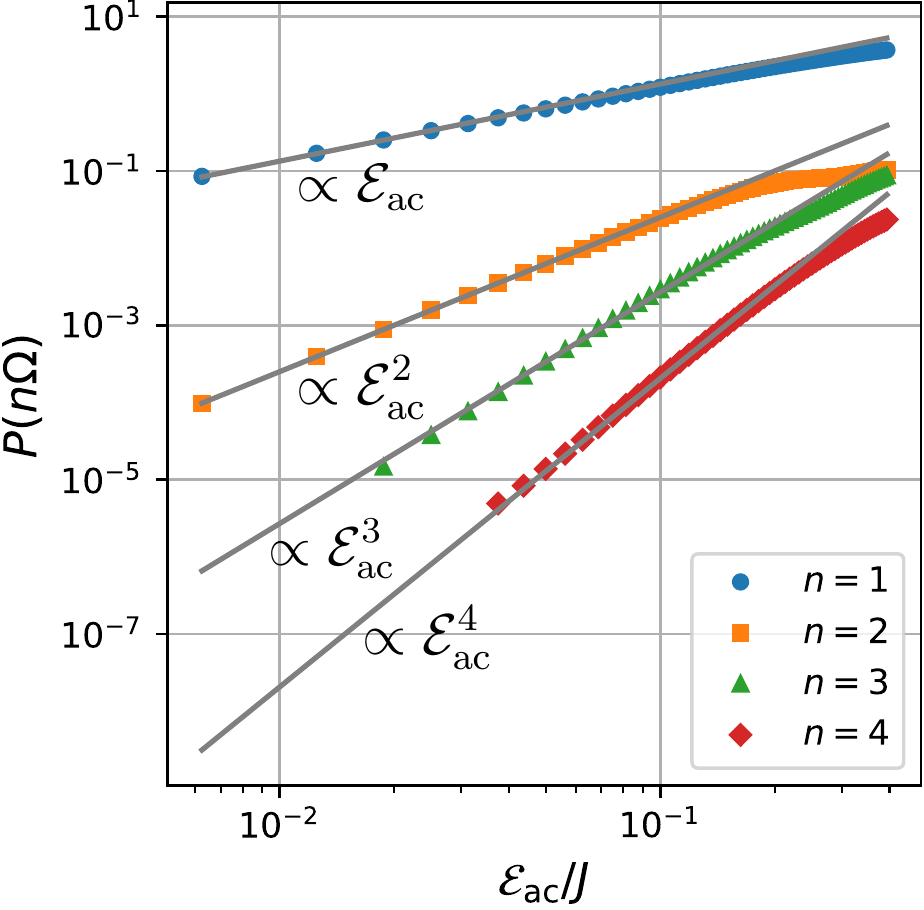}
  \caption{$\eac$ dependence of $n$-th order harmonics of the polarization, $P(n\Omega)\;(n=1,2,3,4)$,
    and their fitting lines ($P(n\Omega)\propto\eac^n$) in the ferromagnetic ($J>0$) Kitaev model with $\edc=0.1J$ and $\kappa=0$
    under THz laser pulse of frequency $\Omega=2.0J$. Due to the small dc electric field of $\edc=0.1J$,
    the even-order harmonics appear. The relaxation rate is set to be $\gamma=0.1J$.}
  \label{fig:depAmplitudeFit}
\end{figure}

First, we consider the crossover.
Figure\autoref{fig:depAmplitudeFit} shows the low-order harmonics of the polarization $P(n\Omega)\;(n=1,2,3,4)$ versus laser intensity in the ferromagnetic ($J>0$) Kitaev model with $\edc=0.1J$ under THz laser with frequency $\Omega=2J$.
When the laser intensity is sufficiently weak, $P(n\Omega)$ increase
with the growth of laser intensity $\eac$. We observe the relation $P(n\Omega)\propto\eac^n$.
This indicates that the perturbation theory is reliable in this weak laser regime.
On the other hand, as the laser intensity is increased,
the power-law relation is gradually violated, and $P(n\Omega)$ becomes smaller
than the fitting curve of $P(n\Omega)\propto\eac^n$.
This violation stems from multiple photon processes
beyond the leading contribution of the perturbation theory.
Figure\autoref{fig:depAmplitudeFit} shows that the deviation from the fitting curve starts
when the laser intensity becomes beyond $\eac/J>0.1$.
Our approach based on Lindblad equation can capture both perturbative and non-perturbative regimes.

\begin{figure}[htb]
  \centering
  \includegraphics[width=0.7\linewidth]{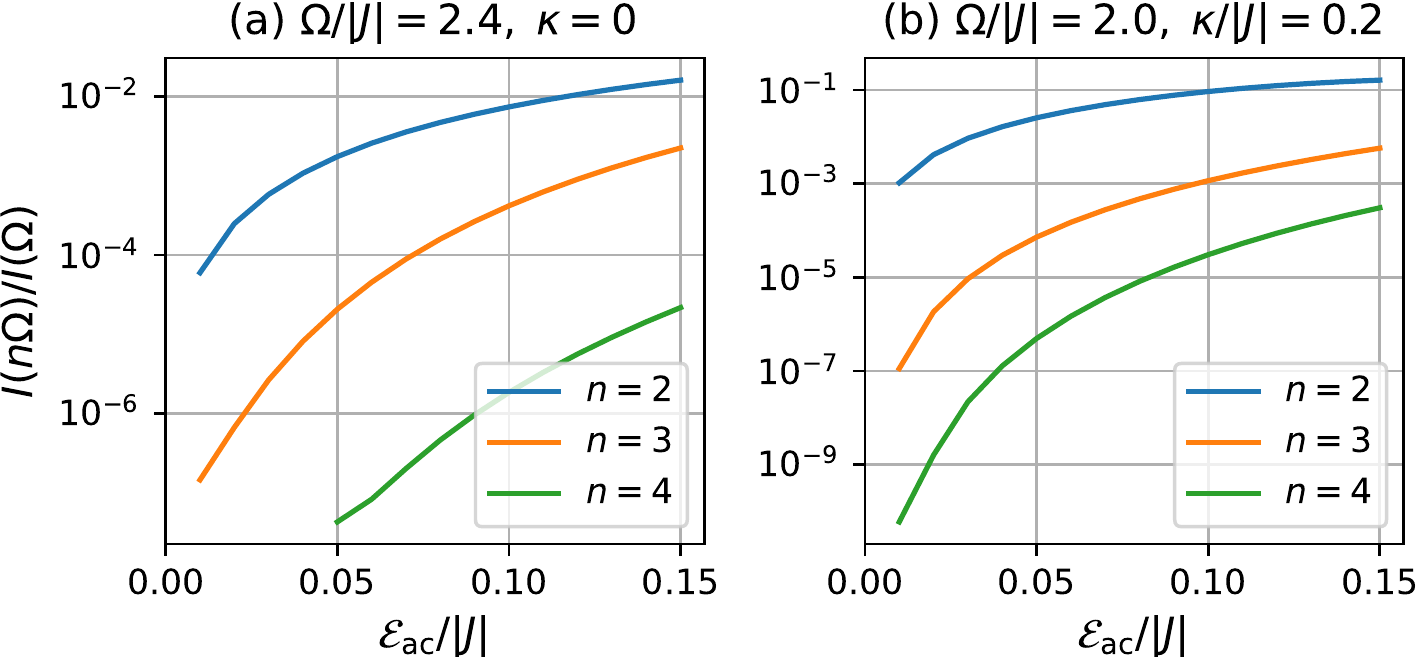}
  \caption{$\eac$ dependence of the intensity rate $I(n\Omega)/I(\Omega)\;(n=1,2,3,4)$
    in the antiferromagnetic ($J<0$) Kitaev models driven by laser pulse of (a) $\Omega=2.4\abs{J}$ at
    $\kappa=0$ and (b) $\Omega=2.0\abs{J}$ at $\kappa=0.2\abs{J}$. Other parameters are set to be $\edc=0.1\abs{J}$ and $\gamma=0.1\abs{J}$.
  }
  \label{fig:depAmplitude234}
\end{figure}

Next, we discuss the laser intensity that is necessary for the experiment.
Figures\autoref{fig:depAmplitude234}(a) and\autoref{fig:depAmplitude234}(b) respectively show the laser-intensity dependence of
the ratio $R_n(\Omega)=I(n\Omega)/I(\Omega)$ at zero magnetic field $\kappa=0$ and
at a finite field $\kappa=0.2\abs{J}$.
As we mentioned in the main text, the peak height of the fermion DoS $\mathcal{D}(\Omega)$ increases
when $\kappa$ is applied to the system. Therefore, for the system with $\kappa=0.2\abs{J}$,
the required laser intensity is expected to become smaller than that for the case of $\kappa=0$.
Comparing Figs.\autoref{fig:depAmplitude234}(a) and\autoref{fig:depAmplitude234}(b),
we find that the ratio $R_{2,3,4}(\Omega)$ become larger when $\kappa$ is introduced in the system.

To estimate the quantitative value of the required laser intensity, we consider the antiferromagnetic
Kitaev magnet with $\abs{J}/k_B=\SI{10}{K}$ and $\edc=0.1J$.
In addition, we assume that the strength of the ME coupling is the same
as that of the standard Zeeman interaction:
$\eac=\eta_{\mathrm{ms}}E_\ac=g_0\mu_{\mathrm{B}}E_\ac/c$ ($c$ is the speed of light).
Under these conditions, one can quantitatively calculate the required laser intensity.
From the panel (a), we see that $E_\ac=\SI{2.6}{MV/cm}$ at \SI{0.5}{THz} is necessary
for $R_2(\Omega)\agt10^{-2}$, and $E_\ac=\SI{0.9}{MV/cm}$ for $R_2(\Omega)\agt10^{-3}$ in the Kitaev model
with $\kappa=0$. On the other hand, Fig.\autoref{fig:depAmplitude234}(b) shows that
in the case of $\kappa=0.2\abs{J}$,
the required ac field $E_\ac$ is estimated as $E_\ac=\SI{0.7}{MV/cm}$ at \SI{0.42}{THz}
for $R_2(\Omega)\agt10^{-2}$, and $E_\ac=\SI{0.2}{MV/cm}$ for $R_2(\Omega)\agt10^{-3}$ (see Table\autoref{tab:necessary}).
From these results, we can conclude that low-order harmonic generation in Kitaev magnets
can be observed with current THz laser technology.

\begin{table}[htbp]
\centering
\caption{Required laser intensity for experimental observation of SHG.}
\begin{tabular}{c @{\hspace{15pt}}| @{\hspace{30pt}} c @{\hspace{30pt}} c}
\hline\hline\rule[-7pt]{0pt}{20pt}
Laser-frequency [$f=\Omega/(2\pi)$]                                         & $R_2(\Omega)\agt10^{-2}$              & $R_2(\Omega)\agt10^{-3}$              \\ \hline\rule[-6pt]{0pt}{20pt}
\SI{0.5}{THz}                        & \SI{2.6}{MV/cm} & \SI{0.9}{MV/cm} \\\rule[-8pt]{0pt}{15pt}
\SI{0.42}{THz} ($\kappa=0.2\abs{J}$) & \SI{0.7}{MV/cm} & \SI{0.2}{MV/cm} \\ \hline\hline
\end{tabular}
\label{tab:necessary}
\end{table}

\end{document}